\documentclass[a4paper,11pt]{article}
\usepackage{jheppub} 

\usepackage{epsfig,multicol,bbm}
\usepackage{enumerate}
\usepackage{bibentry}

\title{General Argyres-Douglas Theory}

\author[a]{Dan Xie}

\affiliation[a]{School of Natural Sciences, Institute for Advanced Study \\
Princeton, NJ 08540, USA}

\abstract{We construct a large class of Argyres-Douglas type theories by compactifying six dimensional $(2,0)$ $A_{N-1}$ theory on 
a Riemann surface with irregular singularities. We give a complete classification for the choices of Riemann surface and the singularities.
The Seiberg-Witten curve and scaling dimensions of 
the operator spectrum are worked out. Three dimensional mirror theory and the central charges $a,c$ are also calculated for some subsets, etc.
Our results greatly enlarge the landscape of $\mathcal{N}=2$ superconformal field theory and in fact also include previous theories constructed using 
regular singularity on the sphere.}

\begin{document} 
\maketitle
\flushbottom

\section{Introduction}
The study of conformal field theory (CFT) plays an important role in
understanding the  dynamics of quantum field theory. The most well known four dimensional
CFT is $\mathcal{N}=4$ supersymmetric gauge theory which has
a conformal manifold labeled by an exact marginal coupling $\tau$:
 gauge coupling constant. 
The weakly coupled description with explicit Lagrangian
description can be written down at the cusp of the coupling constant space. Different weakly coupled descriptions are related by $S$ duality 
which is best understood from the compactification of magical six dimensional 
$(2,0)$ theory  on a torus \cite{Witten:2009at}. 

Recently, a large class of  four dimensional $\mathcal{N}=2$
superconformal field theories (SCFT) are found by compactifying six dimensional $(2,0)$ theory on a Riemann surface with regular punctures \cite{Gaiotto:2009we,Gaiotto:2009hg}. 
Similarly, the gauge coupling constant space is
identified with the moduli space of complex structure of the punctured Riemann
surface and the $S$ duality group is identified with the modular group. However, one usually can not write a Lagrangian description for the weakly coupled 
gauge theory duality frame at the cusps.  The reason is that the matter sectors are usually isolated 
strongly coupled SCFTs which play an essential role in understanding
S duality of these theories, for instance, the S dual theory of $SU(3)$ with
six fundamentals  \cite{Argyres:2007cn} involves strongly coupled $E_6$ theory \cite{Minahan:1996fg} which can be constructed by compactifying six dimensional 
$A_2$ theory on a sphere with three full punctures.

The above theories have dimensionless coupling constants and  the Coulomb branch operators have  integral scaling dimension. 
There are another class of $\mathcal{N}=2$ SCFTs called Argyres-Douglas (AD) theories \cite{Argyres:1995jj} which
 usually have fractional scaling dimensions for the Coulomb branch operators and  dimensional coupling constants. This type of theory is
originally found as the IR theory at special point of Coulomb branch of
pure $SU(3)$ gauge theory. At this special point, mutually nonlocal dyons
become massless and one can not go to a duality frame in which these dyons carry only electric charge. 
The theory must be an interacting SCFT \cite{Argyres:1995jj, Argyres:1995wt} and a Lagrangian description is not possible. There are usually
relevant operators appearing in the spectrum which can be used to deform 
the theories to a new fixed point. 

It is natural to ask whether we can engineer  AD theories using six dimensional $(2,0)$ theory and encode all those physical properties into the geometric objects on the Riemann surface.
The answer is yes and it is necessary to introduce new type of puncture: irregular singularity (higher order pole) \cite{Gaiotto:2009hg}. The previous analysis
 is based on $A_1$ $(2,0)$ theory and we will carry a complete analysis for higher rank theory in this paper.
 It is immediately clear that the possibility for finding new theories are dramatically increased because the order of pole can go to infinity. 
 It is amazingly simple to classify and study these theories once a six dimensional construction is found. 
 In the following, we outline the main strategy of constructing AD theories and summarize the main results.

The Hitchin equation defined on the Riemann surface \cite{Hitchin:1987ab, Hitchin:1987bc} plays a central role in these constructions. Let's first review what happens in the regular puncture case:
The four dimensional UV theory is specified by the choice of Riemann surface with punctures, and the puncture is interpreted as the boundary condition of the fields
in the Hitchin's equation: these punctures have the
regular singularity (first order pole). The gauge coupling constant is identified with the complex
structure moduli of punctured Riemann surface, and the mass parameters are encoded as
the coefficients of the first order pole.
The IR behavior of the four dimensional theory is determined by the moduli space of 
solutions with the above specified boundary condition. In particular, the Seiberg-Witten 
curve is identified with the spectral curve of the Hitchin integrable system.

There is no way to introduce dimensional couplings using only regular punctures 
so the irregular singularity  for the solution of Hitchin's
equation is needed.  The introduction of the irregular singularity provides us the desired properties: first the Coulomb branch operators of 
4d theory can  have fractional scaling dimension as computed from the spectral curve; second the parameters
in the higher order pole are the dimensional coupling constants. The coefficient in the first order 
pole is still identified with the mass parameter, therefore all the deformation parameters 
of the UV theory are matched with the geometric parameters. The moduli space has
similar properties as the regular singularity case, i.e. it is also a hyperkahler manifold \cite{MR2004129}. One can also identify the spectral curve of the 
moduli space with the Seiberg-Witten curve, and the IR behavior of the theory is solved too.

The regular singularity is classified by Young Tableaux and one can put arbitrarily number of 
punctures on a Riemann surface with arbitrary genus to define a SCFT.
The situation is completely different for the irregular singularity case. First, since the coordinate $z$ of
the Riemann surface transforms non-trivially under the 4d $U(1)_R$ symmetry, one can only use
the the Riemann sphere to preserve the $U(1)_R$ symmetry. Second, there are also severe constraints on
the combinations of irregular singularity and regular singularity one can put on a Riemann surface. 
Finally,  the classification of irregular singularity is much more fruitful and not every irregular singularity defines an AD theory. 
To construct an AD theory, we have the following constraints:

1. Only Riemann sphere can be used.

2. There are two singularity combinations: a. Only one irregular singularity; b. One irregular singularity and one 
regular singularity.

3. There are only three classes of irregular singularities.

By specifying the singularity structure on the Riemann sphere, one can write the spectral curve and therefore
find the scaling dimensions of various operators appearing in the theory. The three dimensional mirror
theory for some of the AD theories  can also be determined from the information of the irregular singularity.
 The central charges $a$ and $c$ of those theories can be easily
calculated using 3d mirror theory. The theories we constructed recover almost all the AD theories found in the literature and
there are a lot more new examples.  Not all the SCFTs constructed are distinct and we identify some of the interesting isomorphism which
is useful in calculating the central charge of these theories, etc. 

By allowing the higher order pole, the possible choices for constructing SCFT are greatly increased even with the above constraints.
Our results greatly enlarge the landscape of $\mathcal{N}=2$ SCFT and show that these AD theories are much more generic
than people thought.  In fact, the theory of class ${\cal S}$ (theory using regular singularity) defined on a sphere can also be realized using 
irregular singularity.

Some more properties of these theories are also studied in this paper. First, we use the collision of the singularities to 
identify the AD points of the $SU(N)$ QCD; Second,
if the AD theory has one regular singularity which usually has non-abelian flavor symmetry,  one can use it
to construct new asymptotical free (AF) theory. Geometrically, such AF theories are engineered by putting arbitrary number
of irregular singularities on a Riemann surface.

This paper is organized as follows: in section 2, we give a brief review of AD theory; In section
3, a classification of irregular singularity to Hitchin's equation is given; section 4 discusses the AD theory found from
six dimensional $A_1$ theory, and those theories are not new but we study various properties of these theories  which seem new;
In section 5, AD theories from $A_2$ theory  are classified and studied; section 6
consider the general AD theory found from compactifying six dimensional $A_{k-1}$ theory; Section 7 discuss the 
six dimensional construction of the known AD theories; finally, we give 
a short discussion showing possible further directions in section 8.

\section{Generalities of Argyres-Douglas theory}
The Argyres-Douglas theory is first discovered as the IR theory at certain
point of Coulomb branch of pure SU(3) gauge theory \cite{Argyres:1995jj}. At this point, there
are mutually non-local massless dyons so one can not go to a duality frame
in which all the massless particles carry only electric charge, so a Lagrangian description is not possible. It is further
argued that the theory must be an interacting  SCFT based on the superconformal algebra \cite{Argyres:1995xn}.

Several more examples were also found on the Coulomb branch of $SU(2)$ QCD
by tuning the mass parameters and Coulomb branch parameters \cite{Argyres:1995xn}. For example, the AD theory from $SU(2)$ with only
one flavor is the same as that found from pure SU(3) theory.  The Seiberg-Witten curve of this AD theory is
\begin{equation}
x^2=z^3+mz+u.
\end{equation}
The scaling dimension of the various operators are found by requiring the Seiberg-Witten differential $\lambda=xdz$ to have dimension one:
\begin{equation}
[x]+[z]=1.
\end{equation}
One also require that each term in the Seiberg-Witten curve to have the same dimension, therefore
the scaling dimensions of  coordinates $x$ and $z$ for the above theory are $[x]={3\over 5}$ and $[z]={2\over 5}$. Then it is easy to 
 find the scaling dimensions for the parameters of the theory: $[m]={4\over 5}$ and $[u]={6\over 5}$. 
$u$ is a relevant operator and $m$ is the coupling constant for this deformation. It is
important to have $[m]+[u]=2$ so the above interpretation of $\mathcal{N}=2$ preserving deformation is consistent. 
In fact, for any relevant operator $u_i$ of a AD theory, there has to be a coupling
constant $m_i$ in the spectrum so that $[u_i]+[m_i]=2$ \cite{Argyres:1995xn}. The $U(1)_R$ charge of the
operator is related to the scaling dimension at the superconformal point:
\begin{equation}
R(u_i)=2D(u_i).
\end{equation}
This $U(1)_R$ in the IR has nothing to do with the UV $U(1)_R$ symmetry, though.

It is interesting to note
that the AD theory found in above examples has the same number of parameters as the UV theory, however,
their scaling dimensions are changed dramatically due to the strong quantum effects. In some cases, even the Coulomb branch dimension
is  changed, for example, the above AD theory can also be found from
pure $SU(3)$ gauge theory, and the UV theory has Coulomb branch dimension 2 while
the IR theory has  Coulomb branch dimension one. Similarly, the flavor symmetry 
of the AD theory can be quite different from the UV theory, i.e. the above AD theory has no flavor
symmetry while the model $SU(2)$ with one flavor has a $U(1)$ flavor symmetry.

Let's list some of AD theories found in the literature: many more examples \cite{Eguchi:1996vu,Eguchi:1996ds} are found
on the Coulomb branch of  $SU(N)$ and $SO(N)$ QCD by tuning the parameters, and they found an interesting ADE classification.
The AD theories found from $SU(2)$ QCD  are labelled as $A_0, A_1, A_2$ theory  from singular fibre classification,
and the higher rank generalizations of them are found using F theory by putting multiple D3 branes at the singularity of 
the corresponding type \cite{Banks:1996nj,Douglas:1996js,Gukov:1998kt}.  Recently, a large class of theories based on a pair of Dynkin diagrams are found in \cite{Cecotti:2010fi} using the 2d/4d correspondence.
See also recent investigations of AD theories using F theory in \cite{Seo:2012ns}.

Now let's introduce some of the important quantities associated to a $\mathcal{N}=2$ SCFT.
The four dimensional superconformal field theories have two important central charges
parameterizing the trace anomaly \cite{Kuzenko:1999pi}:
\begin{equation}
<T^\mu_\mu>={c\over 16\pi^2}(Weyl)^2-{a\over 16\pi^2}(Euler),
\end{equation}
where
\begin{eqnarray}
(Weyl)^2=R_{\mu\nu\rho\sigma}^2-2R_{\mu\nu}^2+{1\over3} R^2, \nonumber\\
(Euler)=R_{\mu\nu\rho\sigma}^2-4R_{\mu\nu}^2+R^2.
\end{eqnarray}

For a weakly coupled $N=2$ gauge theory, the central charge can be expressed in terms of
the number of vector multiplets $n_v$ and hypermultiplets $n_h$ \cite{Shapere:2008zf}:
\begin{eqnarray}
c={2n_v+n_h\over 12},~~a={5n_v+n_h\over 24}, \nonumber\\
a-c=\frac{1}{24}(n_v-n_h).
\end{eqnarray}
Here the quantity $a-c$ is proportional to the dimension of ``Higgs" branch in which all the gauge symmetries are completely broken. There is
another useful general equation relating the central charge to the scaling dimensions of the operator spectrum:
\begin{equation}
4(2a-c)=\sum_{i=1}^r(2D(u_i)-1).
\end{equation}
It is usually very hard to calculate the central charges for strongly coupled theories. By
using the topological gauge theory, the author of \cite{Shapere:2008zf} found the following formula:
\begin{equation}
a=\frac{1}{4}R(A)+\frac{1}{6}R(B)+\frac{5}{24}r+{1\over 24}h,~~~~~c=\frac{1}{3}R(B)+\frac{1}{6}r+{1\over 12}h \label{central},
\end{equation}
where $r$ is the number of vector multiplets at the generic point of the Coulomb branch and $h$ is the free
hypermultiplets at generic point; $R(A)$ and $R(B)$ are $R$ charges of certain measure
factors for the topological gauge theory. $R(A)$ can be found from the scaling dimension of the Coulomb branch operators:
\begin{equation}
R(A)=\sum_{i=1}^r(D(u_i)-1),
\end{equation}
and $R(B)$ is related to the discriminant of Seiberg-Witten curve which is in general very difficult to determine. The central 
charges can also be calculated using the supergravity dual \cite{Aharony:2007dj}.

\section{Irregular  singularity of Hitchin's equation}
AD theories are constructed by compactifying six dimensional 
$A_1$ theory on a Riemann surface with irregular singularity \cite{Gaiotto:2009hg}. This description is found 
by taking scaling limits of the Seiberg-Witten curve of known UV theories and then mapping the resulting Seiberg-Witten curve 
to a Hitchin system description.  The generalization of taking scaling limit to higher rank theory is rather difficult and there are 
some subtle points about the scaling limits as discussed recently in \cite{Gaiotto:2010jf}.
Our philosophy is to start directly from the irregular singular solutions of Hitchin's equation, and
use the classification of  irregular singularity to classify 4d theory. The same idea seems working well for the regular 
puncture case in which the solution of the Hitchin equation do have a Young 
Tableaux classification which matches the physical derivation \cite{Nanopoulos:2009uw}.

The new features of the AD theory are the fractional scaling dimension and the dimensional coupling constant, 
and there is no way to accommodate these new features using only regular singularities.  The introduction
of the irregularity automatically solves this problem: first, the coordinate $z$ on the Riemann surface transforms 
nontrivially under $U(1)_R$ symmetry and we have the fractional scaling; second, the coefficients in the higher order pole are the 
dimensional coupling constant. There are many other 
wonderful matchings between the geometric description and the physical quantities as we will discuss
in full detail later.  This section is mainly served as a description of the irregular singularity from the Hitchin equation point of view and 
those who want to see physical applications can skip this section on first reading.

Let's review  various identifications between the physical quantities and the geometric
aspects of Hitchin equation in more detail for the regular puncture cases which are later generalized to the irregular singularity case.
The four dimensional $\mathcal{N}=2$ theory constructed in \cite{Gaiotto:2009we} are derived by compactifying
six dimensional $(2,0)$ SCFT on a Riemann surface with regular punctures (with topological twist). The Hitchin's equation is
also defined on the Riemann surface and the punctures correspond to the regular singular solutions
to Hitchin's equation. The moduli space of solutions to Hitchin's equation
with fixed boundary condition is a hyperkahler manifold and is identified with
the Coulomb branch of four dimensional theory on $R^3\times S^1$, where $S^1$ is a circle.
In fact, the Hitchin moduli space is the Higgs branch of the five dimensional
maximal super Yang-Mills theory compactified on the punctured Riemann surface and
it is the mirror of the original 3d theory \cite{Kapustin:1998xn}. There are no quantum corrections
to the Higgs branch due to the non-renormalization theorem, and that's why the classical picture of the Hitchin equation
encodes the Coulomb branch information of the original theory.

The geometric parameters are the complex structure moduli of the punctured Riemann surface
and the coefficients of the simple pole, which are identified with the gauge coupling constants and mass
parameters respectively. 
The IR behavior of the field theory is encoded into the moduli space which
 is a hyperkahler manifold with  complex structures parametrized by
$CP^1$. In one of complex structure I, the Hitchin's moduli space is an integrable
system and the spectral curve is identified with the Seiberg-Witten
curve. In complex structure J, each point on the moduli space parameterizes
a flat connection on Riemann surface, and this description is useful 
for the classification of the irregular singularity. By taking a complex structure
on the Riemann surface, the Hitchin equation reads
\begin{eqnarray}
F-\phi\wedge \phi=0, \nonumber\\
D\phi=D*\phi=0,
\end{eqnarray}
where $A$ is the connection and $\phi$ is a one form called Higgs field \cite{Hitchin:1987ab,Kapustin:2006pk}.
By writing ${\cal A}=A+i\phi$, the Hitchin equation implies that the curvature of
${\cal A}$ is flat. In fact, one can introduce a spectral parameter and define 
a family of flat connections depending on the spectral parameter. The monodromy of the flat connection ${\cal A}$
around the singularity can be calculated by solving the
following flat section equation:
\begin{equation}
(\partial_z+{\cal A}_z) \psi=0,
\label{flat}
\end{equation}
which locally is just a first order differential equation on the  disk. 

The simplest irregular singular solution to Hitchin's equation with gauge group $SU(N)$ is studied in  \cite{Witten:2007td}, and the fields near the singularity have the form
\begin{eqnarray}
\phi={u_n\over z^n}+.....+{u_2\over z^2}+{u_1\over z}+c.c+....,\nonumber\\
A=\alpha d\theta.
\label{special}
\end{eqnarray}
Here we choose local coordinate $z=re^{i\theta}$, $u_1,...u_n$
are all diagonal matrices with distinct eigenvalues (this can be done by using the gauge symmetry).
 The dot means the  regular terms and we will ignore them in the formula below, but they are always there. 
 This abelianization of the flat connection is crucial for finding the  irregular singular solution to Hitchin's equation.

The Hitchin moduli space in the presence of irregular singularities is also hyperkahler \cite{MR2004129} and they
share many properties as the regular singularity case, in particular, the complex structure I depends linearly
on the coefficient of the first order which should be identified with the mass parameter \cite{Witten:2007td}; Complex structure $I$ 
defines an integrable system and its spectral curve is
\begin{equation}
\det(x-\Phi(z))=0,
\end{equation}
where $\Phi$ is the holomorphic part of $\phi$ and this spectral curve is identified with the Seiberg-Witten curve.
The complex structure J in which each point represents a flat connection is useful for classification and we 
will focus on it in the following. For the irregular singularity presented in (\ref{special}), the homomorphic part of the flat connection  has the following form
\begin{equation}
{\cal A}_z={u_n\over z^n}+{u_{n-1}\over z^{n-1}}+....{u_2\over z^2}+{u_1^{'}\over z}.
\end{equation}

There is an interesting Stokes phenomenon for the differential equation (\ref{flat}) which is important to define the monodromy around the 
irregular singularity.  We will review some aspects for the Stokes phenomenon for the completeness, the interested reader can find
more details in \cite{Witten:2007td,MR919406}. The appearance of 
these  Stokes matrices are coming from the asymptotical behavior of the solutions to the equation $(\partial +{\cal A}_z)\psi=0$.
Let's first assume the gauge group is U(1), then the differential equation  becomes
\begin{equation}
{d\psi\over dz}=-({q_n\over z^n}+{q_{n-1}\over z^{n-1}}...+{q_1\over z}+B(z))\psi,
\end{equation}
here $B(z)$ is a holomorphic function which is regular at $z=0$.  The solution is very simple:
\begin{eqnarray}
\psi=c(z)\exp Q(z),\nonumber\\
Q(z)=({q_n\over(n-1) z^{n-1}}+{q_{n-1}\over (n-2)z^{n-2}}...+q_1(-lnz),\
\end{eqnarray}
here c(z) is a formal power series which is not convergent around the singularity. This solution is the building block for the solution of the higher rank, which just
have a vector of above solution with index $i=1,2,...n$.  However, the entries of solution vector have different asymptotical behaviors 
along different path to the singularity because
\begin{equation}
|{\exp Q^i(z)\over \exp Q^j(z)}|\rightarrow |\exp({q^i_n-q^j_n\over(n-1) z^{n-1}})|,~~~~z\rightarrow 0.
\end{equation}
The asymptotical behavior depends on the sign of $Re({q^i_n-q^j_n\over(n-1) z^{n-1}})$. The sign is different for different angular region and it is easy to 
see that there are $2(n-1)$ angular regions which are called Stokes sector, the boundary of the Stokes sector is called Stokes ray. 

The point is that the solution with given asymptotical behavior in a region is not unique  if there is no stokes ray. For example, if $|{\exp(Q^i(z))\over \exp(Q_j(z)})|>>0$ in this region,
then the solution $\psi_i^{'}(z)=\psi_i(z)+\lambda \psi_j(z)$ has the same asymptotical behavior as $\psi_i$.  Such freedom is not here if there is a Stokes ray in this 
angular region since the asymptotical behavior would be different in two regions separated by the stokes ray, so the solution of two pairs $\psi_i(z)$ and $\psi_j(z)$ in certain region
is fixed if there is a stokes ray in this region.
 
 Let's take a region with angular width $\pi/(n-1)$ whose boundary is not a stokes ray.  By rotating this region by integer value of $\pi\over n-1$, we get a cover of the 
disk.  One can enlarge each sector a little bit such that there is no stokes ray in the overlapping region. 
There will be a stokes ray for any given pair of entries in the solution vector in each sector and the whole solution is uniquely fixed
with given asymptotical behaviors in that region.  On the overlapping region, however, no stokes ray is here, and the two sets of solutions in two regions
are related by an upper triangular matrix with unit diagonal entry.  This matrix is called as the Stokes matrix which is then used to construct the monodromy matrices.

There would be a total of $2(n-1)$ stokes matrices and the product of them defines part of  the generalized monodromy.  The monodromy also has a contribution from
the regular singular term which contribute a $N-1$ parameter. Finally we need to subtract a contribution from the $T_c$ group which has dimension $(N-1)$.
So the total parameters in specifying the local monodromy is
\begin{equation}
 c_n=(n-1)(N^2-N).
\end{equation}
The monodromy  for a path around the irregular singularity with a chosen base point  has another contribution which contributes an extra $(N^2-1)$ parameters.
However, the moduli space is defined by specifying the leading order coefficient which gives a minus $(N-1)$ contribution. So the total contribution to the moduli space
of the irregular singularity is 
\begin{equation}
d=n(N^2-N).
\end{equation}
If there are more than one irregular singularity, the total dimension of the moduli space is the sum of local contribution minus a gauge group contribution
\begin{equation}
dim(M)=\sum_id_i-2(N^2-1).
\end{equation}

The local contribution of the irregular singularity to the moduli space can also be understood in an easy way as the following \cite{MR1904670}: each fixed regular semi-simple matrix defines
a conjugacy glass $O_i$ in the lie algebra. The differential operator $d_{\cal A}$ maps one point from  the product of $n$ conjugacy class to a point on the 
moduli space of flat connections:
\begin{equation}
d_{\cal A}: O_1\times O_2\times O_3\ldots \times O_n\rightarrow M.
\end{equation}
Since each regular semi-simple conjugacy class of $Sl(N)$ has dimension $N^2-N$, the total dimension of the local contribution is $d=n(N^2-N)$ which matches the result
from the consideration of stokes matrices.  

\subsection{Classification of irregular singularity}
Let's give a classification of irregular singularity in this subsection, which is 
achieved by specifying the form of matrices of  the coefficient of higher order pole.  
By introducing $\omega={1\over z}$, the equation (\ref{flat}) \footnote{We do not include the contribution of the gauge fields to the complex flat connection by assuming that
the gauge fields coefficient is in the same conjugacy class as that of first order  coefficient of the Higgs field.} becomes
\begin{eqnarray}
{d\psi\over d\omega}=\Phi^{'}(\omega)\psi, \nonumber\\
\Phi^{'}(\omega)=z^{-2}\Phi(1/z).
\end{eqnarray}
Now the singularity is put at $\omega=\infty$. We put the  singularity at $0$ and $\infty$  interchangeably in this paper, and
the singularity at $\infty$ is good for classification and  finding the Seiberg-Witten curve, while the choice $0$ is good 
for considering the collision of irregular singularity. 

The key point is that the matrices specifying the behavior of the singularity can be put into diagonal form by using formal gauge transformation \cite{Witten:2007td}, and these
diagonal matrices are the ones we want to classify. The leading order matrix  have the following general  form
\begin{equation}
\Phi(z)= diag(z^{r_1-2}B_1, z^{r_2-2}B_2,\ldots, z^{r_s-2}B_s),
\end{equation}
here $B_i$ is a $k_i\times k_i$ diagonal matrix whose eigenvalue degeneracy will be discussed later, and $r_1> r_2 > r_3\ldots >r_s$ are 
a set of rational numbers denoting the order of  pole of various blocks. 
The crucial point now is that the resulting spectral curve should have only integer power in $z$, therefore the rational number $r_i$ should have the following form
\begin{equation}
r_i=n_i+{j_i\over k_i}, ~~~0< j_i\leq k_i,
\end{equation}
with $n_i, j_i, k_i$ all positive integers.
Now let's study the degeneracy of eigenvalues of various matrices and we look at $B_1$ block only (other blocks are similar).

If $j_1\neq k_1$, the eigenvalues can only have the following form (assume $j_1$ and $k_1$ have no common divisor, the degenerating case is considered later):
\begin{equation}
B_1=diag(1,\omega, \omega^2, \ldots, \omega^{k_1-1}), ~~~j_1\neq k_1
\end{equation}
with $\omega=\exp({2\pi i\over k_1})$ (we do not write this value explicitly  in later sections, but $\omega$ always have 
the above form with proper choice of $k_1$).  This specific form is derived by requiring the spectral curve to have the integer pole on the coordinate $z$.
The meaning of such fractional power 
in Higgs field  is explained in \cite{Witten:2007td}: there is a cut coming out  of
singularity and a gauge transformation is done to put the gauge field $\Phi(\theta+2\pi)$ back to 
$\Phi(\theta)$ \footnote{ One can also do a gauge transformation to put the 
above solution into a form with integer pole, but it is not good for our purpose and we will stick to the above exotic representation.}.
The gauge transformation is basically a permutation on the eigenvalues, since by circling around the singularity, the new solution is
\begin{equation}
B_1^{'}=diag(\omega^{j_1},\omega^{j_1+1}, \omega^{j_1+2}, \ldots, \omega^{j_1+k_1-1}), ~~~j_1\neq k_1.
\end{equation}
A permutation  on the eigenvalues implemented by the gauge transformation would lead the Higgs field back to the original form .
The sub-leading matrices are determined by the requirement that every possible deformation 
compatible with leading order matrix is allowed: the same permutation we introduced earlier
should transform it to its original form after circling around the singularity,  so the following terms
should be allowed:
\begin{equation}
z^{m+{l\over k_1}-2}(\omega^d,\omega^{d+1}, \omega^{d+2}, \ldots, \omega^{(k_1+d-1)}),
\end{equation}
where $m, l$ are  integers and $-1\leq(m+{l\over k_1})<r_1$, and $d$ is a fixed integer depending only on $l$.
 Moreover, the following  terms are also allowed
\begin{equation}
z^{m-2}(a,a,\ldots, a),
\end{equation}
with $m<r_1$  an integer, since apparently any permutation would lead back to the original form.

If $j_1=k_1$,  the pole $r_i$ is integer and the eigenvalues of $B_i$ can take any degenerating form
\begin{equation}
B_1=diag(a,\ldots,a, b,\ldots, b, c,\ldots c,\ldots),~~~~j_1=k_1,
\end{equation}
and the degeneracy can be recorded using a Young Tableaux $Y$ with total boxes $k_1$, i.e. the height of 
first column is the number of eigenvalues with value $a$, and so on. The sub-leading terms also have 
the integer pole and the degeneracy of the eigenvalues are encoded by a larger Young tableaux. For example,  if the 
leading order Young Tableaux is $[3,2,1]$ which means that the first three eigenvalues are the same, and 
then second two eigenvalues are the same, etc; The sub-leading term can be represented by a Young Tableaux $[2,2,1,1]$, namely
now the first three eigenvalues are decomposed into two blocks: two of them are still the same, but the third one is now 
different.

\begin{figure}[htbp]
\small
\centering
\includegraphics[width=8cm]{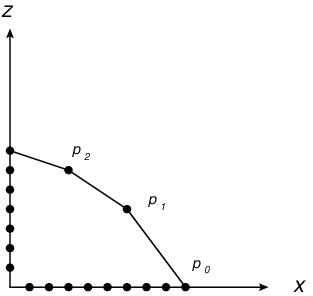}
\caption{The Newton polygon representing an irregular singularity, and each segment on the boundary represents a block and its slope is
the order of pole of that block.}
\label{convex}
\end{figure}

There is a nice graphic representation for the leading order matrices which turns out to be really useful to find the Seiberg-Witten curve.
 The irregular singularity is represented by a convex Newton polygon on a two dimensional integer lattice.
 The convex polygon is determined as follows: first find a point  $p_1$ \footnote{The $x$ coordinate is required to be less than $N$, and the $z$ coordinate is positive.}
   such that  the line between $p_1$ and point $p_0=(N,0)$ has slope $r_1-2$. 
 Then  another point $p_2$ \footnote{The $x$ coordinate of $p_i$ is smaller than $p_{i-1}$, and the $z$ coordinate of $p_i$ is bigger than $p_{i-1}$.} is chosen such that the slope of the line $p_2p_1$ has the value ${r_2-2}$, etc. 
 We have a convex polygon at the end. See figure.~ \ref{convex}.
The above graph is not enough to determine the irregular singularity if there is a segment with integer slope, one also need to specify the degeneracy of the matrices of 
the lower order pole.  Here are some important examples:

a. The irregular singularity considered in last subsection corresponds to $r_1=n$ with $n$  an integer, and there is only one block.  
Moreover, the eigenvalues of $B_1$ are all distinct.

b. There is also only one block, but now the leading order of pole $r_1$ can be fractional.

c. The leading order has two blocks, and it takes the following form:
\begin{equation}
\Phi=z^{r-2}diag(0,1,\omega,\ldots,\omega^{N-2}).
\end{equation}

d. The leading order have only one block and  $r_1=n$ is an integer. The matrix $B_1$ is specified by a Young Tableaux $Y_n$ which represents the 
degeneracy of eigenvalues of $B_1$. Then we need a collection of Young Tableaux such that $Y_n\subseteq Y_{n-1}\ldots\subseteq Y_1$,
where $Y_{j-1}$ is derived by further partitioning each column of  $Y_{j}$. 

\begin{figure}[htbp]
\small
\centering
\includegraphics[width=10cm]{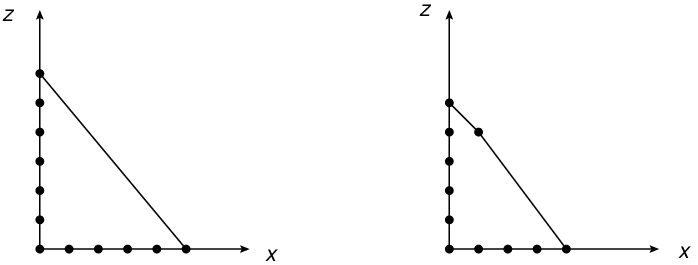}
\caption{Left: The Newton polygon for the irregular singularity with only one block and this is the case $a$ and $c$. Right: The leading order has partition $[N-1, 1]$, this is the case $b$.}
\label{irr2}
\end{figure}

Notice that the four  classes have one thing in common: the leading order matrix has essentially only one block, and those 
irregular singularities are the one needed for defining the AD theories.
Moreover, the leading order matrix of case $a$, $b$, $c$ all have distinct eigenvalues. 
The case $a,b$ can be represented by a Newton polygon as shown in the left of figure. \ref{irr2} with appropriate 
dots on the boundary for $a$, and case $c$ is represented on the right of figure.\ref{irr2}.

The local dimension of the irregular singularity can also be found by counting the deformation parameters  which keep the above form of the 
irregular singularly, i.e. the eigenvalue degeneracies are not changed. For example,
if the singularity is defined using a sequence of Young Tableaux (case d), then one just count the dimension of the corresponding adjoint orbit where 
the number has been given in \cite{Gukov:2006jk}.  The Coulomb branch dimension is equal to half of the moduli space. 
One can also count the dimension of local moduli space by studying the Stokes matrices which will appear elsewhere \cite{Xie:2012jd}, 
and such Stokes matrices are important in finding the corresponding cluster coordinates for 
the field theory \cite{Xie:2012jd}.  The number of  mass parameters are read from the form of the matrices of the first order pole. 
The number of coupling constant is found from the number of parameters of the irregular singularity in the diagonal form. 

\newpage
\section{AD points from 6d $A_1$ theory}
Let's start with six dimensional $A_1$ theory and compactify it on a Riemann surface with irregular punctures. Although we did not find any 
new theories, it is helpful to see some generic features of our construction which 
is then generalized to higher rank. Moreover, with our current construction, we could find their three dimensional mirror, the central charges $a$ and 
$c$, and their $SU(2)$ linear quiver UV completions, etc. 

We need to first figure out what kind of 
singularity combinations are allowed to have a SCFT in the IR.
The four dimensional $\mathcal{N}=2$
SCFT has a $U(1)_R$ symmetry which has a geometric meaning in the six dimensional construction: it is
the rotational symmetry of the Riemann surface and so coordinates $z$ and $x$ of cotangent bundle transform under this symmetry.
 In the case of regular singularity, the $z$ coordinate
on the Riemann surface transform trivially under this symmetry, so one can
put arbitrary number of regular singularities on Riemann surface with any
genus. $z$ coordinate transforms non-trivially (if order of pole $r>2$) for the irregular singularity case, so
the singularity should be put at the fixed points of the rotational symmetry. It is
well known a $U(1)$ isometry with fixed points can only be found for the Riemann sphere, and in this
case, the fixed points can be put at the south pole or north pole. So one can
put at most two singularities to construct AD theory.

There are two kinds of irregular singularities for SU(2) Hitchin system
from our classification. The form of the holomorphic Higgs field (the singularity is at $\infty$) is 
\begin{equation}
\Phi={\lambda z^{n-2}}\left[\begin{array}{cc}
1&0\\
0&-1
\end{array}\right]+{A_{n-1}} z^{n-3}+....+{A_1\over z}+\ldots,
\end{equation}
 we call this type I singularity. The other solution is
\begin{equation}
\Phi={\lambda z^{n-5/2}}\left[\begin{array}{cc}
1&0\\
0&-1
\end{array}\right]+{A_{n-1} z^{n-7/2}}+....+{A_2\over z^{1/2}}+.....
\end{equation}
There is a cut in $z$ plane, in crossing the cut, a gauge transformation is needed
to make the solution consistent.  Notice that there
is no first order term and therefore no mass parameter is encoded in this singularity.
We call this type II singularity.

We now argue that only one irregular singularity is allowed, and
one can only turn on a regular singularity if a irregular singularity is already here . The proof is following:
If we put two irregular singularities at $z=0,\infty$, the spectral curve has the form
\begin{equation}
x^2=z^N+\ldots+{\lambda^2\over z^M}.
\end{equation}
Here $M\geq 3$ and $\lambda$ is the coefficient from higher order pole. We also use the scale transformation to set the coefficient of
$z^N$ to be 1. If this defines a SCFT, then $[\lambda^2]={2(N+M)\over N+2}>2$ which means it is
a Coulomb branch operator, but this is in contradiction to the fact that the parameters from
the coefficient of higher order pole should be the coupling constant which must have dimension less than one.
In summary, we have the following singularity configurations which will define an AD theory:

1. One irregular singularity at the Riemann sphere.

2. One irregular singularity at south pole, and another regular singularity at the north pole of sphere.

\subsection{The construction of AD points}
\subsubsection{One irregular singularity: $(A_1, A_{N-1})$ SCFT}
The four dimensional theory is defined by putting one irregular singularity on the infinity of
the Riemann sphere.  Let's first describe the number of coupling constants, mass parameters and 
the dimension of the Coulomb branch from the geometric data.

The local dimension of type I singularity to Hitchin's moduli space is $2n$. 
There are a total of $n-2$ parameters in $A_{n-1},\ldots,A_2$ and a  mass parameter is encoded in $A_1$. The parameter 
in $A_{n-1}$ can be eliminated using translation invariance, so effectively there are $n-3$ coupling constants.  The Hitchin moduli space has dimension $(2n-6)$ by including a global contribution
 and the dimension of the base of Hitchin fibration is $n-3$ which equal to the dimension of the Coulomb branch, see table. \ref{T1}

The local dimension of type II singularity to Hitchin's moduli space is also $2n$, and we still have $n-3$ parameters in $A_{n-2},..A_2$ (again we use translation invariance to eliminate 
the parameter in $A_{n-1}$), but
there is no mass parameter. The  base of Hitchin's 
fibration also has dimension $n-3$, see table. \ref{T1}.
\begin{table}[htbp]
    \begin{tabular}{|c|c|c|c|c|}
        \hline
        ~ & Order of pole &Base dimension & First order & Higher order \\ \hline
        Type I&n & n-3 & 1&n-3 \\  \hline
        Type II&n-1/2& n-3 & 0 &n-3\\ 
        \hline
    \end{tabular}
    \caption{The geometric data for one irregular singularity on Riemann sphere, one parameter in higher order pole is eliminated using the translation invariance.}
    \label{T1}
\end{table}

The Seiberg-Witten curve is derived by calculating the spectral curve of Hitchin's fibration
\begin{equation}
x^2=Tr(\Phi^2)=z^N+u_2z^{N-2}+.....+u_N.
\end{equation}
It is easy to see $N=2n-4$ for type I singularity and $N=2n-5$ for type II singularity.  This class of theories are called $(A_1, A_{N-1})$ theory which comes from the fact
that the BPS quiver for this theory is of the product of $A_1$ and $A_{N-1}$ Dynkin diagram as shown in \cite{Cecotti:2010fi,Cecotti:2011rv,Gaiotto:2009hg}. In the following, the label of theory by a pair of Lie 
algebra always means that the BPS quiver  has the shape of product of two Dynkin diagrams.

\begin{figure}[htbp]
\small
\centering
\includegraphics[width=10cm]{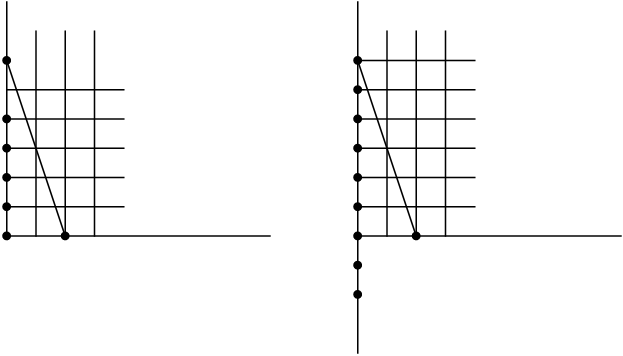}
\caption{Left: A graph representation of $(A_1, A_5)$ theory, scale invariance is used to fix the  coefficient of $z^6$ term to be 1 and translation invariance is used
to eliminate $z^5$ term; Right: A graph representation for $(A_1, D_7)$ theory, notice that $z^5$ term is turned on as the coupling constant.}
\label{AD1}
\end{figure}

The Seiberg-Witten curve can be nicely read from the Newton polygon of the irregular singularity. 
The non-negative points bounded by the Newton polygon represent the monomial appearing in the Seiberg-Witten curve: each
lattice point with coordinate $(m,n)$ represents a monomial $x^mz^n$.
The monomial with one $x$ factor is missing because $\Phi$ is traceless and
One use the translation invariance  to eliminate the points on the line $z=N-1$, see left graph on figure.\ref{AD1}.

The Seiberg-Witten differential is $\lambda=xdz$, and
the scaling dimensions for $x$  and $z$ are
\begin{equation}
[x]={N\over N+2},~~~[z]={2\over N+2}.
\end{equation}
Using this, one can calculate the scaling dimension
of various operators from the Seiberg-Witten curve :
\begin{equation}
D(u_i)={2i\over N+2}.
\end{equation}

If $N=2n-5$, $u_2$ to $u_{n-2}$ have scaling dimension less than one and $u_{n-1}$ to $u_{2n-5}$ are 
the relevant operators, so the Coulomb branch dimension is $n-3$ and the number of  coupling constants are 
$n-3$.   For each relevant operator $u_i$, there is a
coupling constant $m_i$ in the spectrum such that $D(u_i)+D(m_i)=2$.

Similarly, if $N=2n-4$, $u_2$ to $u_{n-2}$ are coupling constants, and $u_{n-1}$ has dimension one which is 
the mass parameter. $u_{n}$ to $u_{2n-4}$ are coulomb branch parameters with a total number
of $n-3$. The data is summarized in table. \ref{T2}.
\begin{table}[htbp]
    \begin{tabular}{|c|c|c|c|c|c|c|cl}
        \hline
        ~  & Coulomb branch  & Mass parameter & Coupling constant &Singularity \\ \hline
        $(A_1,A_{2n-5})$ & n-3 & 1&n-3 &Type I with order n \\  \hline
     $(A_, A_{2n-6})$ & n-3 & 0 &n-3 &Type I with order n-1/2 \\
        \hline
    \end{tabular}
    \caption{The counting of physical parameters from Seiberg-Witten curve .}
    \label{T2}
\end{table}

By comparing Table. \ref{T1} and Table. \ref{T2}, we find that the number of physical quantities from Seiberg-Witten curve 
are exactly the same as the prediction from the geometric data (considering the translation and scaling invariance).

In fact, $(A_1, A_2)$ theory is the original AD theory found from $SU(2)$ with one flavor, and
$(A_1, A_3)$ is the AD theory found from $SU(2)$ with two flavors. 

\subsubsection{One irregular singularity, One regular singularity: $(A_1, D_{N+2})$ SCFT}
We could  add one more regular singularity at point $0$  to define another class of SCFT. Geometrically, this regular singularity 
introduces a new mass parameter and a $SU(2)$ flavor symmetry;  moreover the Coulomb branch dimension is increased by one.
The Seiberg-Witten curve is
\begin{equation}
x^2=Tr(\Phi^2)=z^N+u_1z^{N-1}+.....+u_N+{u_{N+1}\over z}+{m^2\over z^2}.
\end{equation}
This time the translation invariance is fixed by two punctures and $u_1$ can not be eliminated,
so we have a new coupling constant $u_1$, which pairs with the relevant operator $u_{N+1}$. 
$m^2$  has dimension two and represents the mass term from the regular singularity.
This class of theory is called $(A_1, D_{N+2})$ theory as the BPS quiver has the corresponding
shape. In fact, when $N=-1$, it is nothing (no fundamental), and it represents one fundamental of $SU(2)$ when $N=0$
as shown in \cite{Gaiotto:2009hg,Nanopoulos:2010zb}. The first nontrivial theory is $(A_1, D_3)$ theory
which is also the AD points found from $SU(2)$ with two flavors, which is natural since $D_3$ and $A_3$ have the same Dynkin diagram. $(A_1, D_4)$ theory is the 
AD theory found from  tuning mass parameters and Coulomb branch operators of $SU(2)$ gauge  theory  with 3 flavors..

The BPS spectrum and wall crossing behavior of above two class of theories are studied in \cite{Shapere:1999xr, Gaiotto:2009hg,Alim:2011ae,Alim:2011kw} , 
and extended object like line operators and surface operators are studied in \cite{Gaiotto:2010be,Gaiotto:2011tf}.

\subsection{Three dimensional mirror}
In the case of regular singularities, one can compactify four dimensional theory
on a circle and then flow to the deep IR to get an interacting three dimensional
${\mathcal N}=4$ SCFT $A$. This theory has Coulomb branch and Higgs branch. There is a 
mirror theory $B$ for which the Higgs branch of $B$ is the Coulomb branch of $A$ and vice versa.
It is amazing that the mirror theory always has a Lagrangian description \cite{Benini:2010uu} : a star-shaped quiver.

\begin{figure}[htbp]
\begin{center}
\includegraphics[width=10cm]{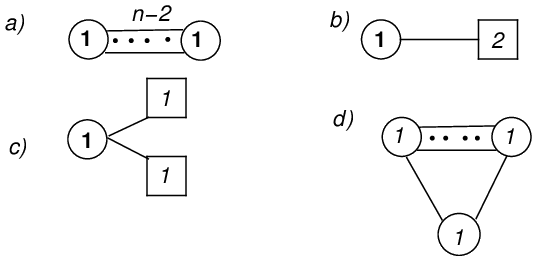}
\end{center}
\caption{(a): Quiver for type I irregular singularity with order $n$, there are $n-2$
bifundamentals between two $U(1)$ groups, this is the 3d mirror for $(A_1, A_{N-1})$ theory, here $N=2n-4$;  b): Quiver leg
for a regular singularity; c): We spray the $U(2)$ flavor symmetry of $(b)$ to two $U(1)$s so that we
can glue this tail to  the quiver in $(a)$; d): Gluing quiver tail in $c$ and the quiver in $a$ which is the three dimensional mirror for $(A_, D_{N+2})$ theory.}
\label{AD2}
\end{figure}

Similarly, Compactifying the four dimensional AD theory on a circle and flow to the deep IR,
we should get a three dimensional ${\mathcal N}=4$ SCFT too. In fact, the mirror theory can also be found by
gluing the quiver components of each singularity \cite{Boalch:2008pb,Nanopoulos:2010bv}.
The rule is the following:

a. Attach a quiver leg as shown in figure. \ref{AD2}b for each regular singularity.

b. Attach a quiver for type I irregular singularity as shown in  figure.~ \ref{AD2}a.

We spray the $U(2)$ node of the regular singularity into two $U(1)$ nodes as shown in figure. \ref{AD2}c. The gluing 
 is achieved by identifying the U(1) nodes as shown in figure. \ref{AD2}c.
We can see the enhanced flavor symmetry of the original theory from the symmetry on the Coulomb branch of the mirror theory.
For example, $(A_1, A_3)$ theory has $SU(2)$ flavor symmetry which can be seen from the 3d mirror. The 3d mirror of this 
theory has two $U(1)$ gauge group and 2 bi-fundamentals between them, since one of the $U(1)$ is decoupled, so the final
mirror theory is $U(1)$ with two flavors which is just the $T(SU(2))$ theory and it is well known that the symmetry on the Coulomb branch 
is $SU(2)$.

\newpage

\subsection{AD points from linear quiver}
In original paper \cite{Argyres:1995jj,Argyres:1995xn}, AD points are found from
Coulomb branch of $N=2$ $SU(2)$ SQCD by tuning the parameters. It would be desirable to find 
a similar UV theory for all the AD theory found on previous section. Instead of taking various scaling
limit of the original theory, we look at the singularity structure  needed for engineering these theories by using Hitchin system.
The irregular singularity for the SQCD has lower order pole and the corresponding AD theory has higher 
order pole, so irregular singularity for the AD theory could be derived by colliding the lower order 
singularity of the corresponding SQCD. It is necessary to find a rule for colliding singularity though.

The  six dimensional constructions of various $SU(2)$ SQCD is worked out in \cite{Cherkis:2000ft, Gaiotto:2009hg}. By comparing
the irregular singularity of these SQCDs and the corresponding AD theories, it is easy to guess the general rule. See fig.~\ref{SU(2)}.
For example, the $SU(2)$ with one flavor has a type I singularity with $n=2$ and a type II singularity also with $n=2$. The corresponding AD theory
has only one  type II irregular singularity with $n=4$ which could be thought of as colliding two order 2 irregular singularities.  
The crucial point is that the number of parameters encoded in the new irregular singularity should be the same as the sum of the 
original two. Since we assume all the parameters (mass, Coulomb branch parameters) are the parameters (mass, coupling constant, Coulomb branch parameter)
of the AD theory. By analyzing the other cases in figure. \ref{SU(2)},  we find the following rules:

\begin{figure}[htbp]
\small
\centering
\includegraphics[width=10cm]{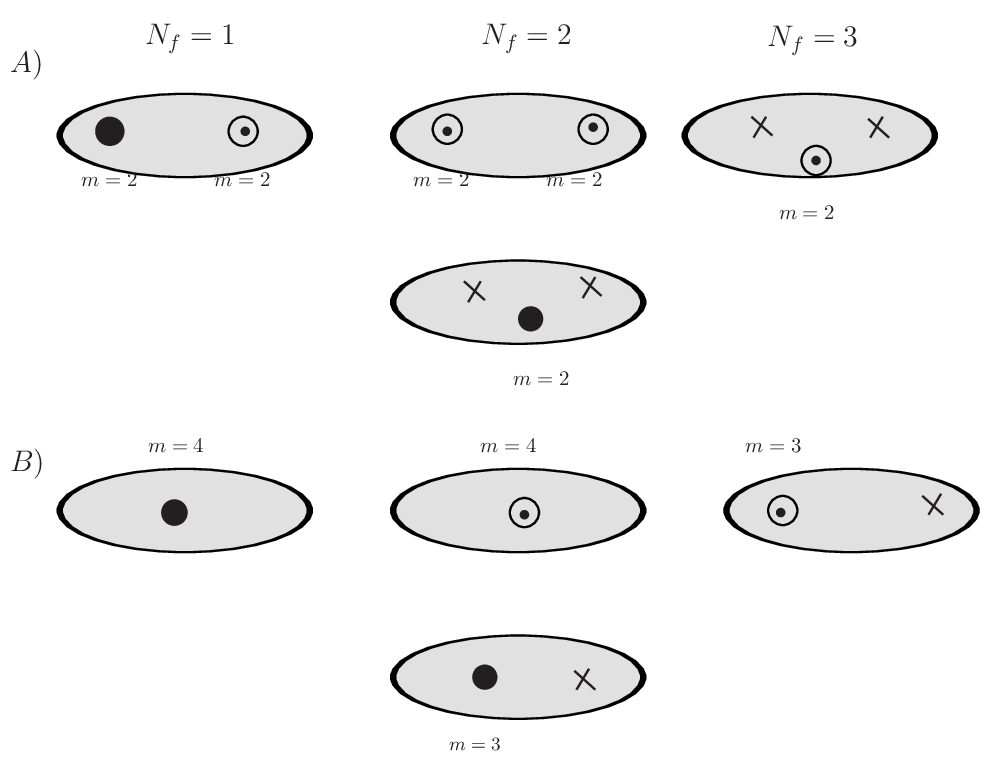}
\caption{A) The Hitchin system for various SU(2) QCDs, black dots represent type II irregular
singularity;  Circle represents  type I irregular singularity;
Cross represents the regular singularity. The number is the integer part of the order of pole. 
 B) The singularity structure for the corresponding AD theory found from the above QCD.
}
\label{SU(2)}
\end{figure}

If we have at least three singularities, one can only do the following collision

a. Colliding a irregular singularity with a regular singularity, and the order of pole of the irregular singularity is increased by one and the type is the same.

If we have two irregular singularities left, we can 

b. Colliding two irregular singularities with order $n$ and $m$, the new irregular singularity has order $n+m$.

Using the above rule, it is not hard to find the UV theory of the AD theories constructed before.
There are four AD theories with $(n-1)$ dimensional Coulomb branch: $(A_1, A_{2n-1})$, $(A_1, A_{2n-2})$, $(A_1, D_{2n})$, $(A_1, D_{2n-1})$.
According to above rule, one can identify one class of the UV theory. The UV theory is the $SU(2)-SU(2)$ linear quiver with different number of
fundamentals on both ends.

1.  $(A_1, A_{2n-2})$ $\rightarrow$ $0-\underbrace{SU(2)-...-SU(2)}_{n-1}-1$.

2.  $(A_1, A_{2n-1})$ $\rightarrow$ $1-\underbrace{SU(2)-...-SU(2)}_{n-1}-1$.

3.  $(A_1, D_{2n-1})$ $\rightarrow$ $0-\underbrace{SU(2)-...-SU(2)}_{n-1}-2$.

4.  $(A_1, D_{2n})$ $\rightarrow$ $1-\underbrace{SU(2)-...-SU(2)}_{n-1}-2$.

Let's explain the colliding rule of case 1 and other theories can be understood similarly.  The linear quiver theory has an order $2$  type I singularity, an order $(2-1/2)$ type II singularity and 
$(n-2)$ regular singularities. One first collide type I singularity with the regular singularities to produce an order $n$  type I singularity, and at the 
end collide with the type II singularity to produce a higher order $(n+3/2)$ type II singularity which is the one for the $(A_1, A_{2n-2})$ theory. 
The UV theory has a total of $(2n-2)$ parameters and the IR theory also have $(2n-2)$ parameters.

\subsection{Central Charge $a$ and $c$}
There are two methods to calculate the central charges $a$ and $c$ of the AD theories constructed in this section.
The first one is to use the  following two formulas 
\begin{eqnarray}
(2a-c)={1\over 4}\sum_{i=1}^r(2D(u_i)-1),\nonumber\\
a-c={1\over 24}(n_v-n_h).
\end{eqnarray}
The second equation is valid for the weakly coupled free theories, we assume that
this is also true for strongly coupled theory if we regard $n_v$ and $n_h$ as the effective number of vector multiplets
and hypermultiplets. Notice that $n_v-n_h$ is just  minus the dimension of ``Higgs" branch of
the SCFT. This number is equal to the dimension of  Coulomb
branch of  the mirror which is easy to calculate because the mirror has a Lagrangian description.  The Coulomb branch dimension of the 3d mirror  is $1$ for 
$(A_1, A_{N-1})$ theory for $N=2n$. Let's first apply the above formulas to $(A_1, A_{N-1})$ theory and get
\begin{eqnarray}
2a-c={1\over 4}\sum_{i=n+2}^{2n}({4i\over 2n+2}-1)={1\over 4}{(-1 + n) (1 + 2 n)\over(1 + n)}, \nonumber\\
a-c=-{1\over 24}.
\end{eqnarray}
Solving above equations, we get
\begin{equation}
a={-5 - 5 n + 12 n^2\over 24 (1 + n)},~~c={3n^2-n-1\over 6(n+1)}.
\end{equation}

For $(A_1, D_{N+2})$ theory and $N=2n$, we could do the similar calculation using the fact that the Coulomb branch dimension of the 3d mirror is 2:
\begin{eqnarray}
2a-c={1\over 4}\sum_{i=n+2}^{2n+1}({4i\over 2n+2}-1)={n\over2}, \nonumber\\
a-c=-{1\over 12}.
\end{eqnarray}
Solving it and we get the central charges
\begin{equation}
a=\frac{n}{2}+\frac{1}{12},~~~~c=\frac{n}{2}+\frac{1}{6}.
\end{equation}

To calculate the central charges for $N=2n-1$, we use the following formula derived from topological field theory (\ref{central}):
\begin{equation}
a=\frac{1}{4}R(A)+\frac{1}{6}R(B)+\frac{5}{24}r,~~~~~c=\frac{1}{3}R(B)+\frac{1}{6}r,
\end{equation}
where $r$ is the dimension of the Coulomb branch and $h$ is zero for the AD theory, and
\begin{equation}
R(A)=\sum_i(D(u_i)-1),
\end{equation}
here the summation is over all the Coulomb branch operators, which is easy to calculate using the explicit Seiberg-Witten curve.
 $R(B)$ is related with the discriminant of Seiberg-Witten curve which is in general very hard to calculate.  The idea is to assume that 
$R(B)$ is a universal function of $N$  for each class of theories, and use the above result for $N=2n$ to find the function dependence of 
 $R(B)$ on N.

For $(A_1, A_{N-1})$ SCFT with $N=2n$, use the scaling dimensions of the operators from Seiberg-Witten curve, we have (see also \cite{Gaiotto:2010we})
\begin{equation}
N=2n:~~~R(A)={n(n-1)\over 2(n+1)}. 
\end{equation}
Substitute the known results of central charges $a$ and $c$ in to the formula (\ref{central}), then $R(B)$ has the following form
\begin{equation}
(A_1, A_{N-1}): R(B)={N(N-1)\over 2(N+2)}. 
\end{equation}

Similarly, for the $(A_1, D_N)$ theory,  it is easy to calculate
\begin{equation}
N=2n:~~~R(A)={n\over2},
\end{equation}
and find the following result for  $R(B)$:
\begin{equation}
(A_1, D_{N+2}):~~~R(B)={N+1\over 2}.
\end{equation}

So finally, we can calculate the central charges for other theories which we do not have the 3d mirror construction, but we use the explicit
form of $R(B)$ nose. The central charges for the AD theory when $N=2n-1$ is 
\begin{eqnarray}
(A_1, A_{N-1}):~~~a={(n-1)(24n-5)\over 24(2n+1)},~~~c={(n-1)(6n-1)\over 6(2n+1)}. \nonumber\\
(A_1, D_{N+2}):~~~a={n(8n+3)\over 8(2n+1)},~~~c={n\over 2}.
\end{eqnarray}
Such results are in perfect agreement with the results in \cite{Shapere:2008zf, Gaiotto:2010we}. A consistency check is that 
both $(A_1, D_3)$ and $(A_1,A_3)$ theory give the same answer $a={11\over 24}$ and $c={1\over 2}$. The central charge formula is summarized in table. \ref{central}.

\begin{table}[htbp]
\begin{center}
    \begin{tabular}{|c|c|c|c|}
        \hline
        ~  & Constraint&a  &c \\ \hline
        $(A_1,A_{N-1})$ & $N=2n$& ${-5 - 5 n + 12 n^2\over 24 (1 + n)}$ & ${3n^2-n-1\over 6(n+1)}$ \\  \hline
     $(A_1, A_{N-1})$ & $N=2n-1$ &${(n-1)(24n-5)\over 24(2n+1)}$ & ${(n-1)(6n-1)\over 6(2n+1)}$  \\  \hline
     $(A_1, D_{N+2})$ & $N=2n$ & $\frac{n}{2}+\frac{1}{12}$ & $\frac{n}{2}+\frac{1}{6}$ \\ \hline
      $(A_1, D_{N+2})$ & $N=2n-1$ & ${n(8n+3)\over 8(2n+1)}$ & ${n\over 2}$ \\
        \hline
    \end{tabular}
    \caption{The central charges of AD theory from six dimensional $A_1$ $(2,0)$ theory.}
    \label{central}
    \end{center}
\end{table}

\subsection{More singularities: Gauge theory coupled with $(A_1,D_{N+2})$ theory}

\begin{figure}[htbp]
\begin{center}
\includegraphics[width=4in]{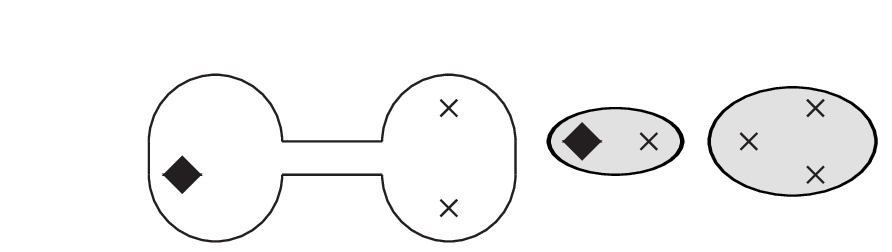}
\end{center}
\caption{(a)A Riemann sphere with one irregular punctures and two regular punctures. (b) In its
``degeneration limit'', the matter are two fundamentals and an AD point we defined earlier.
}
\label{AF1}
\end{figure}

We can put arbitrary number of irregular singularities and regular singularities on
any Riemann surface.  These theories are asymptotical free theories which can be confirmed  
using the contribution to the $\beta$ function of the AD theory if the non-abelian flavor symmetry is gauged \cite{Gaiotto:2010jf}.
Physically, the matter parts are the three sphere with regular punctures which represents 
the tri-fundamental  and the sphere with one irregular singularity and one regular singularity representing AD theory.
The full theory is derived by gauging the diagonal $SU(2)$  flavor symmetry of the regular singularity.
See figure. \ref{AF1} for an example. 

\begin{figure}[htbp]
\begin{center}
\includegraphics[width=1.0in]{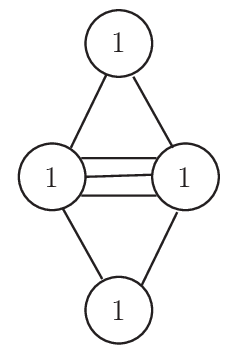}
\end{center}
\caption{The three dimensional mirror for the gauge theory described in the last figure.
}
\label{ADmirror}
\end{figure}

 In the case of sphere with just one type I irregular singularity and several 
regular singularities, one can find its 3d mirror using the prescription described earlier, one 
example is shown in figure. \ref{ADmirror}. Those theories are called complete theories and 
their BPS spectrum is studied in \cite{Cecotti:2011rv}, in particular, the BPS quiver of those theories 
are of the finite mutation type \cite{fomin-2008-201}.

\newpage
\section{AD points from 6d $A_2$ theory}
Let's now start with six dimensional $A_2$ theory and compactify it on a Riemann surface with irregular singularity.
We would like to find a four dimensional AD theory in the IR.
The analysis for the type of Riemann surface and the number of irregular singularities is completely the same
as for the $SU(2)$ case by focusing the degree two differential in the Seiberg-Witten curve. We can only 
use Riemann sphere and there can be either one irregular singularity or one irregular singularity plus a regular 
singularity. There are more choices for irregular singularity for $A_2$ theory while 
 only two type of irregular singularities for $A_1$ theory exist, so immediately we find many more  new AD type 
 theories.

\subsection{Classification of irregular singularity for AD theories}
\begin{figure}[htbp]
\begin{center}
\includegraphics[width=10cm]{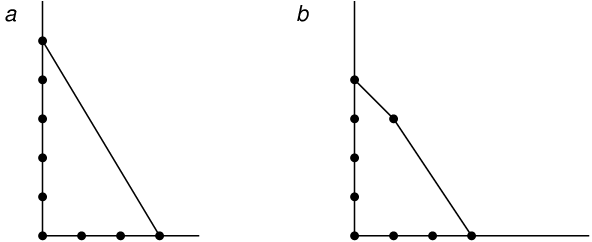}
\end{center}
\caption{a: The Newton polygon for the type I irregular singularity. b: Newton polygon for type II irregular singularity. }
\label{S3}
\end{figure}

Based on our  classification for irregular singularities, we have the following
catalog of irregular singularities  for $SU(3)$ group:

Type I:  The order of pole for the irregular singularity and the leading order matrix are
\begin{eqnarray}
\Phi=z^{r-2}diag(1,\omega,\omega^2), \nonumber\\
r=n+j/3,~~0<j\leq 3.
\end{eqnarray}
For the integer pole, the leading order matrices does not necessarily have the above specified form, and the diagonal terms can be three 
arbitrary numbers. There are also two mass parameters encoded as the coefficient  of the first order pole in this case.
There are no mass parameter if the order of pole is fractional.  The number of coupling constants (exclude the leading order 
and the first order coefficients) are  found by counting all the possible deformation compatible with the leading order matrix:
\begin{eqnarray}
N_{coupling}=2n+j-3, ~~j\neq 3, \nonumber\\
N_{coupling}=2(n-1),~~~j= 3.
\end{eqnarray}
The Newton Polygon for this type of irregular singularity 
is depicted in figure. \ref{S3}a.

Type II: The order of pole for the irregular singularity and the leading order matrix are
\begin{eqnarray}
\Phi=z^{r-2}diag(0,1,\omega), \nonumber\\
r=n+1/2.
\end{eqnarray}
There is one mass parameter from the coefficient of the first odder pole and the number of coupling 
constants are:
\begin{equation}
N_{coupling}=2(n-1).
\end{equation}
The Newton Polygon for this type of irregular singularity 
is depicted in figure. \ref{S3}b. 

Type III: One can consider the degeneration of the irregular singularity with integer order of pole $n$. The irregular singularity is 
now labeled by two integers $(n_1,n_2)$ such that $n_1+n_2=n$, namely, there are $n_1$ simple Young Tableaux and 
$n_2$ full Young Tableaux. The number of mass parameters are determined by the Young Tableaux $Y_1$: there is one if
$Y_1$ is simple and two if $Y_1$ is full. The number of coupling constants are
\begin{equation}
N_{coupling}=n_1+2n_2-3.
\end{equation}

The above information is summarized in table. \ref{T3}.

\begin{table}[htbp]
    \begin{tabular}{|c|c|c|c|c|c|c|}
        \hline
        ~  & Order & Base dimension & First order & Higher order \\ \hline
       Type I & $n+j/3$ & 3n+j-7 & 0&2n-3+j \\  \hline
       Type I & $n+1$& 3n-5    &2&2(n-1)+one marginal\\ \hline
    Type II &$n+1/2$ &${3\over2}(n-1)$  &1 & 2(n-1) \\ \hline
    Type III&$n=n_1+n_2$ &$2n_1+n_2-8$ &2 or 1&$n_1+2n_2-3$ \\ \hline
    \end{tabular}
    \caption{The counting of parameters from the geometric data, here we count the maximal possible number of parameters in higher order coefficients of irregular singularity.}
    \label{T3}
\end{table}

\subsubsection{Type I SCFT: $(A_2, A_{N-1})$ theory}
Let's compactify six dimensional $A_2$ theory on  on Riemann sphere with first type of irregular singularity, and 
we will get a four dimensional AD theory in the IR.
The Seiberg-Witten curve can be easily found from the spectral curve  using 
the form of Higgs field on the puncture.  In practice, it is actually much easier to read it  directly from the Newton polygon of the corresponding irregular singularity. 
The lattice points bounded by the newton polygon represent the allowed monomials appearing in the Seiberg-Witten curve.
We use the scale invariance to set the coefficient of the $z^N$ term to be 1 and use the translation invariance to 
eliminate the $z^{N-1}$ term. The points on $x=2$ line  are not used because the trace of the Higgs field is zero.
\begin{figure}[htbp]
\begin{center}
\includegraphics[width=10cm]{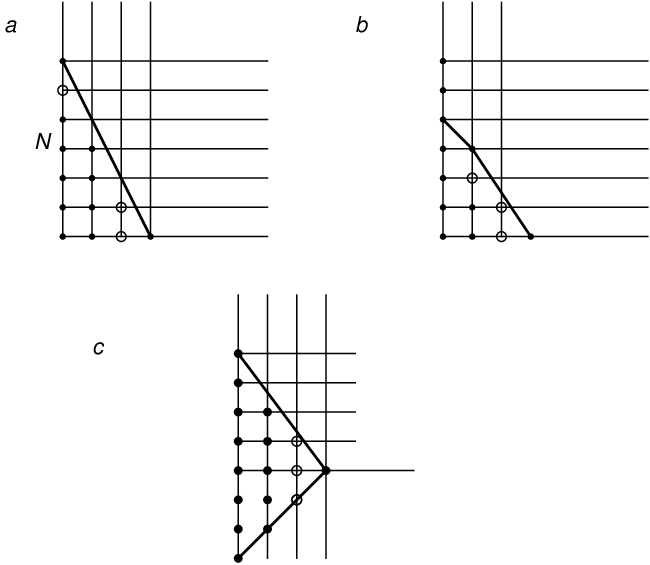}
\end{center}
\caption{The Seiberg Witten curve can be read from the integer points bounded by the Newton polygon. For each 
included point with coordinate $(m,n)$, there is a corresponding monomial $x^mz^n$ in the Seiberg Witten-curve.}
\label{S33}
\end{figure}

The  $(A_2, A_{N-1})$ theory is described by the Seiberg-Witten curve $x^3+z^{N}=0$, and the scaling dimension 
for the coordinates are determined by requiring the Seiberg-Witten differential $\lambda=xdz$ to have dimension one:
\begin{equation}
[x]={N\over N+3},~~~[z]={3\over N+3}.
\end{equation}
The SW curve under general deformations are  found by using the bounded lattice points of the Newton polygon. 
 Let's take $N=3n-1$ for an  detailed example, other cases are similar. The Seiberg-Witten curve is
\begin{equation}
x^3+(v_1z^{2n-1}+\ldots+v_iz^{2n-i}+\ldots+v_{2n})x+(z^{3n-1}+u_1z^{3n-3}+\ldots+u_{3n-2})=0.
\end{equation}
One can find the scaling dimension of the various operators appearing in the Seiberg-Witten curve using the scaling dimension of the coordinates,:
\begin{equation}
[v_i]={3i-2\over 3n+2},~~~ [u_i]={3+3i\over 3n+2}.
\end{equation}
It is easy to see that for each relevant operator  $O_i$ in the spectrum, there is another coupling constant $m_i$
such that $D(O_i)+D(m_i)=2$. There is no mass parameter in the spectrum and  the Coulomb branch dimension 
is $3n-2$; there are also  $2n$  operators with dimension less than one, and they are the coupling constants for the relevant deformations.

Let's compare the above numbers with the parameters in the definition of  irregular singularity.  The order of pole for the irregular singularity  
 is 
 \begin{equation}
 r=(3n-1)/3+2=(n+1)+2/3.
\end{equation}
Since the order of pole is fractional, there is no mass parameter which matches very well with the result from the SW curve. The total
number of coupling constants from the irregular singularity are $ 2n+1$. Subtracting one coupling constant using the translation invariance, the final number matches the 
result from Seiberg-Witten curve. One can also check that the dimension of the Coulomb branch is the same 
as the dimension of the base of the Hitchin fibration. 

We should point out that $(A_2, A_3)$ theory is equivalent to $(A_1, E_6)$ theory as discussed also in \cite{Cecotti:2010fi}. 
The Seiberg-Witten curve at the AD point is
\begin{equation}
x^3+z^4=0.
\end{equation}
and the scaling dimension of all the operators has common denominator 7. The $(A_2, A_4)$ theory is isomorphic to
$(A_1, E_8)$ theory since the SW curve at the fixed point is 
\begin{equation}
x^3+z^5=0.
\end{equation}
and the common denominator of the scaling dimension is $8$ which is in agreement with the result in \cite{Cecotti:2010fi}.  Using 
our method, we can construct all the deformations for these two theories, moreover, it is easy to construct the 
BPS quiver directly using the information in irregular singularity, and they do has the same form as the corresponding 
Dynkin diagram  \cite{Xie:2012jd}.

\subsubsection{Type II SCFT}
This class of theories are constructed using the second type of irregular singularity. 
The Seiberg-Witten curve is also easily found from the Newton polygon from figure. \ref{S3}b, with $N=2n-1$:
\begin{equation}
x^3+(z^{2n-1}+v_1z^{2n-3}+\ldots +v_iz^{2n-2-i}+\ldots+v_{2n-2})x+(u_1z^{{3n-2}}+\ldots+u_iz^{3n-1-i}+\ldots+u_{3n-1})=0.
\end{equation}
Now the scaling dimension is determined by the singularity $x^3+z^{2n-1}x=0$ and by requiring the Seiberg-Witten differential $\lambda=xdz$ to have the 
dimension one, we have
\begin{equation}
[x]={2n-1\over 2n+1},~~~[z]={2\over 2n+1}.
\end{equation}
The scaling dimension for all the parameters appearing in Seiberg-Witten curve can be easily found:
\begin{equation}
[v_i]={2i+2\over 2n+1},~~~~[u_i]={2i-1\over 2n+1}.
\end{equation}
One can check that there is a coupling constant $m_i$ for each relevant operator $u_i$ such that
$D(m_i)+D(u_i)=2$.  There is one mass parameter in the spectrum which is in agreement with the result in table. \ref{T3}.
The order of pole of the irregular singularity is 
\begin{equation}
r={2n-1\over 2}+2=(n+1)+{1\over2}, 
\end{equation}
so there are $2n$ coupling constants in the irregular singularity from table. \ref{T3},  and one of them can be eliminated using translation 
invariance, so we have $2n-1$ parameters from the geometry which matches perfectly with the number of 
coupling constants read from the Seiberg-Witten curve.

When $N=3$, the AD theory is equivalent to $(A_1,E_7)$ theory. The
SW curve at the singularity is $x^3+xz^3=0$ and the common denominator for the scaling dimension is
5 which is in agreement as what is discovered in \cite{Cecotti:2010fi}. 
The BPS quiver  of this theory  indeed has the shape of $E_7$ Dynkin diagram as will be analyzed in \cite{Xie:2012jd}. Using our construction, 
we identify all the deformations for this theory.

\subsubsection{Type III SCFT}
This class of theories are defined using the the class 3 singularity which has integer order of pole $n$ and the first $n_1$ matrices  
of the irregular singularity have the partition $[2,1]$ while the last $n_2$ coefficients have 
partition $[1,1,1]$. The Seiberg-Witten curve is the same as the the one with leading order
singularity regular semi-simple, but not all the operators in the Seiberg-Witten curve are independent. 

The Seiberg-Witten curve has the following form
\begin{equation}
x^3+\phi_2(z)x+\phi_3(z)=0.
\end{equation}
We would like to calculate the maximal order $d_i$ of $z$ in $\phi_i(z)$ whose coefficient gives the independent Coulomb branch operator,
and they can be read from the Young Tableaux. For a single Young Tableaux, the contribution to $\phi_i(z)$
\begin{equation}
p_i=i-s_i,
\end{equation}
where $s_i$ is the height of $i$th box in the Young Tableaux. For the partition $[2,1]$, we
have $p_2=1, p_3=1$, and for the partition $[1,1,1]$, the order of pole is $p_2=1, p_3=2$.
The leading order with coefficient representing real Coulomb branch parameters is determined by
\begin{equation}
d_i=\sum_j p_i^{(j)}-2i,
\end{equation}
and the summation is taken over all the Young Tableaux in the definition of the irregular singularity, so $d_2=n-4,~~d_3=n_1+2n_2-6$ and the Coulomb branch dimension is 
\begin{equation}
d=d_2+d_3+2=2n_1+3n_2-8=2n+n_2-8.
\end{equation}

The SW curve at the AD point is $x^3+z^{3n-6}=0$ and the scaling dimension of $[x]$ and $[z]$ is 
\begin{equation}
[x]={n-2\over n-1},~~[z]={1\over n-1}, 
\end{equation}
the minimal scaling dimensions of the Coulomb branch operators appearing in $\phi_2$ and $\phi_3$ are 
\begin{equation}
[u_1]=2[x]-d_2[z]={n\over n-1},~~[v_1]=3[x]-d_3[z]={2n-n_2\over n-1}.
\end{equation}
The number of relevant operators from $\phi_2$ are therefore $(n-3)$ while
the relevant operators from $\phi_3$ are $n_2-2$.  The total number of relevant operators are
\begin{equation}
N_{relevant}=n-3+n_2-2=n_1+2n_2-5.
\end{equation}
The number of parameters in the definition of the irregular singularity is $n_1+2n_2-3$, so we do have enough 
coupling constants from the irregular singularity and two of them is frozen though.

However, not all of those theories are SCFT. The reason is that there is no exact marginal deformation from the geometric data, but 
the Coulomb branch spectrum can consist of operator with scaling dimension two. Our interpretation is that these theories are asymptotical free
theory. The theory without  dimensional two operator are $n_2=1$ or $n_2=0$, and they give new SCFT.

\subsubsection{Type IV: One regular singularity, One irregular singularity}
We can add one more regular singularity at point zero to previous one irregular singularity example. 
Let's just take $(A_1,A_{3n-1})$ as an example,  other cases are just completely  similar.
When adding one more full regular singularity, the Seiberg-Witten curve is 
\begin{eqnarray}
x^3+(v_1z^{2n-1}+....v_iz^{2n-i}+v_{2n}+{v_{2n+1}\over z}+{v_{2n+2}\over z^2})x+\nonumber\\
(z^{3n}+\omega z^{3n-2}+u_1z^{3n-2}+....+u_{3n-1}+
{u_{3n}\over z}+{u_{3n+1}\over z^2}+{u_{3n+2}\over z^3})=0.
\end{eqnarray}
Notice that there is a new term $\omega$ which is forbidden in the one singularity case.
There are three more Coulomb branch parameters $v_{2n+1}$, $u_{3n}$, $u_{3n+1}$, with dimension
\begin{equation}
[v_{2n+1}]={6n+3\over 3n+3},~~~[u_{3n}]={9n+3\over 3n+3},~~~~[u_{3n+1}]={9n+6\over n-1}.
\end{equation}
And $v$ is always a relevant operator.  $\omega$ is a coupling constant with scaling dimension
\begin{equation}
[\omega]={3\over 3n+3}.
\end{equation}
So $\omega$ and $v_{2n+1}$ matches well, the other two parameters $v_{2n+2}$ and $u_{3n+2}$ are
just mass parameters with dimension $2$ and $3$ respectively. 

If the regular singularity is simple, only $v_{2n+1}$ and $u_{3n}$ are the independent Coulomb branch operator,
also there is only one new mass parameter.

\subsection{Three dimensional mirror theory}
\begin{figure}[htbp]
\begin{center}
\includegraphics[width=10cm]{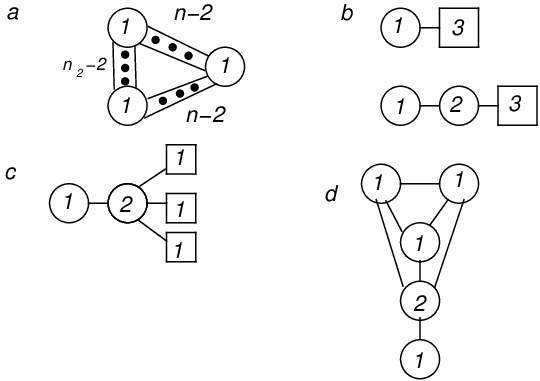}
\end{center}
\caption{a: The quiver for the irregular singularity with $n_1$ minimal Young Tableaux and $n_2$ full 
Young Tableaux, and the order of pole is $n=n_1+n_2$. b:  The quiver tail for the simple regular puncture and the full regular puncture. c: the U(3) flavor
symmetry of the regular singularity quiver tail is split into three U(1) factor. d: Glue the sprayed quiver tail of 
the regular singularity to the irregular singularity and form the 3d mirror theory; there is an order 3 irregular 
singularity with leading order regular semi-simple, and one more full regular singularity.}
\label{s3mirror}
\end{figure}

If we compactify four dimensional theory on a circle and flow to deep IR, the IR theory is a three dimensional
$\mathcal{N}=4$ SCFT $A$. It is straightforward to find the mirror theory $B$ \cite{Intriligator:1996ex} from the irregular singularity if
the order of pole is integer.  Assume that the first  $n_1$ partitions has Young Tableaux $[2,1]$ and the last
$n_2$ partitions has Young Tableaux $[1,1,1]$, then one can associate a quiver as shown in
figure. \ref{s3mirror}.   If $n_2=0$, one have only two nodes whose ranks are one and two, there are
$n-2$ arrows between these two nodes, in this case, the number of mass parameters of the original theory 
is one.

There are some easy checks: the Higgs branch dimension of this quiver is $2n+n_2-8$ which is exactly 
the Coulomb branch of the original theory. The number of FI parameters are just the number of quiver nodes
minus one which match the number of  mass parameters of the original theory.

The quiver tail for the regular singularity is worked out in \cite{Benini:2010uu} and is completely fixed 
by the Young Tableaux, see figure. \ref{s3mirror}. To glue the quiver tail of 
the regular singularity to the quiver of the irregular singularity, we need to spray the $U(3)$ flavor
symmetry of the regular singularity according to the quiver of the irregular singularity, i.e. if there are
three $U(1)$ nodes for the irregular singularity, then we spray the $U(3)$ flavor symmetry of the
regular singularity into three $U(1)$s. The gluing is achieved by identifying these $U(1)$ flavor
symmetries with the $U(1)$ gauge group of the quiver for irregular singularity.

\newpage

\subsection{Central charges $a$ and $c$}
The strategy of finding the central charges $a$ and $c$ is quite the same as we have done for 
$SU(2)$ theory: one can first calculate the central charges using the three dimensional mirror 
symmetry and then using the result to find a universal function $R(B)$ for that class; finally, 
$R(B)$ is used to calculate the central charges for other theories in this class.

Let's first consider the $(A_2, A_{N-1})$ theory with $N=3n$ since the 3d mirror theory is known. 
The Seiberg-Witten curve using the Newton polygon has the following form:
\begin{equation}
x^3+(v_1z^{2n-1}+....v_iz^{2n-i}+v_{2n})x+(z^{3n}+u_1z^{3n-2}+....+u_{3n-1})=0.
\end{equation}
The scaling dimensions of $x$ and $z$ are
\begin{equation}
[x]={3n\over 3n+3},~~[z]={3\over 3n+3}.
\end{equation}
Using this information, one can find the scaling dimensions of the operators:
\begin{equation}
[v_i]={3i\over 3n+3},~~~[u_i]={3i+3\over 3n+3}.
\end{equation}

The Coulomb branch of the mirror is just $2$ and so the effective Higgs branch dimension of 
the original theory is also $2$, using the following formula,   
\begin{eqnarray}
2a-c={1\over 4}\sum_{i=1}^r(2D(u_i)-1),\nonumber\\
a-c=-{1\over 12}.
\end{eqnarray}
we find the central charges:
\begin{eqnarray}
a={1\over 4}\sum_{i=1}^r(2D(u_i)-1)+{1\over 12}={-5 - 5 n + 24 n^2\over12 (1 + n)}, \nonumber\\
c={1\over 4}\sum_{i=1}^r(2D(u_i)-1)+{1\over 6}=-{1 + n - 6 n^2\over3 + 3 n}.
\end{eqnarray}
Notice the summation is taken over all the operators with dimension larger than one.

Now we would like to calculate $R(B)$ using the above result and the formula (\ref{central}). $R(A)$ is easily found from the scaling dimensions of the spectrum
\begin{equation}
R(A)=\sum_{i=n+2}^{2n}[{3i\over 3n+3}-1]+\sum_{i=n+1}^{3n-1}[{3i+3\over 3n+3-1}]={n (-3 + 5 n)\over2 (1 + n)}.
\end{equation}
Substitute the above results of the central charges into the formula (\ref{central}), one have 
\begin{equation}
R(B)={3N(N-1)\over 2(N+3)}.
\end{equation}

We assume $R(B)$ having the universal form and can be applied to other theories in this class, then the 
central charges for them can be calculated easily.
Similarly, one could find the central charges for other SCFT if there is an explicit 3d mirror theory. This includes the type III SCFT and 
type IV SCFT in which the irregular singularity has 3d mirror quiver. We leave this to the interested reader.

\subsection{AD theories from $SU(3)$ QCD}
We now use the collision of the singularity idea to find the possible AD locus of the $SU(3)$ QCD. The 
irregular singularity types of a certain $N_f$ theory is not unique and depend on a partition of $N_f=n_1+n_2$
with $n_i\leq 3$ \cite{Nanopoulos:2010zb}, this is coming from putting different number of branes on left
and right hand side \cite{Witten:1997sc}. The singularity types depend on the number $n$ in the following way:

a. $n=0$, $\Phi=\lambda z^{-1-1/3}diag(1,\omega, \omega^2)+\ldots$. This is a type I irregular singularity.

b. $n=1$, $\Phi=\lambda z^{-1-1/2}diag(0, 1, \omega)+\ldots$. This is a type II irregular singularity

c. $n=2$, $\Phi=\lambda z^{-2}diag(-2,1,1)+z^{-1}(m_1,m_2,m_3)+\ldots$. This is a type III irregular singularity.

d. $n=3$, there are a full regular singularity and a simple regular singularity.

The rules for collision can be found by requiring the combined irregular singularity has the same number of 
parameters as the original one. Here are the rules:

1. In the fractional pole case, one can only collide two irregular singularities which are of the same type, 
or collide the fractional irregular singularity with the integral irregular singularity whose leading order
matrix is regular semi-simple.

2.  One can also collide the irregular singularity with fractional order with the full regular singularity.

3. The collision of two irregular singularities are allowed if they are the only singularities left;
The collision of the irregular singularity with the regular singularity is allowed if there are at least three singularities.

In all the cases, the order of pole of combining singularity is simply the sum of original two.
Let's having some fun using the above rules. 

For $N_f=0$, there are 
two identical type I irregular singularities with pole order $1+1/3$, the collision will produce a type I irregular singularity with order of pole 
$r=2+2/3$.  The AD theory from this irregular singularity is $(A_2, A_1)$ theory. The AD theory found in  \cite{Eguchi:1996vu} is 
actually $(A_1, A_2)$ theory, which are indeed identical as we show later.   There are two parameters in the UV which are
just the Coulomb branch operators; for the corresponding AD theory, one also have two operators: a relevant operator and 
a coupling constant. So the UV parameters and IR parameters match.

For $N_f=2$, there 
are two type II irregular singularities if we put one brane on left and right side. The collision will produce a type I singularity with $r=3$. The AD points 
corresponding this singularity are in fact as  that same found in \cite{Eguchi:1996vu}, which is actually the $(A_2, A_2)$ theory.
The UV theory has four parameters: two mass parameters and two Coulomb branch operators. The AD theory also have two mass parameter, and one 
Coulomb branch operator and one coupling constant. 

For $N_f=3$, one put all three branes on one side, then there would be a type I irregular singularity, a full regular singularity and a simple 
regular singularity. The collision between the irregular singularity and the full regular singularity produce a type I irregular singularity with $r=2+1/3$. 
Our AD theory is described by a type I irregular singularity with pole order $r=2+1/3$ and a simple regular singularity. The 
AD theory is a type IV theory with 5 parameters which match the UV theory.

For $N_f=4$, the splitting is $4=3+1$. There would be a type II irregular singularity, a full regular singularity and a simple regular singularity. By colliding 
the irregular singularity and the full regular singularity, one find a type IV AD theory with 6 parameters.

We are not able to find any AD theory for the $N_f=5$ theory from colliding irregular singularity. 

\subsection{The use of the type IV SCFT}
When there are more than one irregular singularities on sphere, the four dimensional theory is an asymptotical free theory.  
The new matter content appearing in the ``degeneration" limit is the type IV SCFT which is represented by two punctured 
sphere.  For example, if there are two irregular singularities, then the four dimensional theory is a $SU(3)$ gauge group coupled with two type IV SCFTs
defined using the corresponding full regular singularity and the corresponding irregular singularity.   The physical picture 
is the same as depicted in figure. \ref{AF1}.

\begin{figure}[htbp]
\begin{center}
\includegraphics[width=8cm]{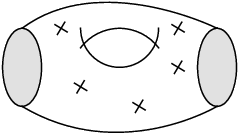}
\end{center}
\caption{Riemann surface with a bunch of regular and irregular singularities, and the 4d theory is an asymptotical free theory which is formed by gauging Type IV
Argyres-Douglas theory and the SCFT formed by three punctured sphere.}
\label{AF3}
\end{figure}

With the above observation, there is no problem of writing the matter contents and weakly coupled gauge group (including the asymptotical gauge group)
of the corresponding four dimensional theory for any combinations of irregular singularities and regular singularities on 
a genus $g$ Riemann surface.  Geometrically, all the irregular singularity should be on the boundary of the Riemann surface and the 
regular singularity is sitting on bulk.  See figure. \ref{AF3} for an example, each boundary represents an irregular singularity, physically, this represents 
a type IV AD theory coupled with the bulk.

\section{AD points from 6d $A_{k-1}$ theory}
\subsection{The choices of irregular singularity}
Now let's start with a six dimensional $A_{n-1}$ theory and compactify it on a Riemann surface with irregular singularity. 
First we would like to determine what kind of irregular singularity is needed for defining a 4d SCFT. The analysis 
for the number of irregular singularity is the same as the previous section: one could have only one irregular 
singularity, or one irregular singularity and a regular singularity on north and south pole on the sphere.  In $A_1$ and $A_2$ 
case, all kinds of irregular singularities define SCFT in 4d.  The situation is different for higher rank group: not every irregular singularity defines a 
4d AD theory.

The key is that the parameters from the higher order pole should be the coupling constants, so
the operators in the Seiberg-Witten curve should have scaling dimension less than one if the operators are formed only by 
those parameters, i.e those terms do not contain parts from the regular terms of the Higgs field. 
 This condition puts severe constraints on the type of irregular singularity one can use to find a SCFT.

The definition of the irregular singularity depends on a sequence of slopes which can be used to draw a 
Newton polygon. The sequence is arranged such that $\lambda_1> \lambda_2> \ldots> \lambda_l$, here $\lambda_i=r_i-2$ where
$r_i$ is the maximal order of pole of $i$th block in irregular singularity.
Let's first  take $\lambda_1=n_1+j_1/k_1$ with $j_1<k_2$ and $n_1\geq 2$, and the
 leading order matrix of this block has the following form (the singularity is put at $z=\infty$):
\begin{equation}
B_1= z^{n_1-2+{j_1\over k_1}}diag(1,\omega,....\omega^{k_1-1}).
\label{canonical}
\end{equation}
This  block determines the scaling dimensions of the various operators and the Seiberg-Witten curve at the singularity takes the form
\begin{equation}
x^n+x^{n-k_1}z^{k_1n_1-2k_1+j_1}=0,
\end{equation}
so $z$ and $x$ have the scaling dimension
\begin{equation}
[z]={k_1\over k_1n_1-k_1+j_1}~ ,~~[x]={k_1n_1-2k_1+j_1\over k_1n_1-k_1+j_1}.
\label{scaling}
\end{equation}

Now let's turn on the deformation corresponding to second block with dimension $k_2>1$ and order of 
pole of this block is  $r_2=(n_1+j_2/k_2)<r_1$ (we take $n_2=n_1$ so that $[v]$ can have smallest possible scaling dimension), and $j_2/k_2<j_1/k_1$. The SW curve looks like:
\begin{equation}
x^n+x^{n-k_1}z^{k_1n_1-2k_1+j_1}+x^{n-(k_1+k_2)}z^{k_1n_1-2k_1+j_1}(z^{k_2n_1-2k_2+j_2}+\ldots+vz^{(k_2-1)(n_1-2)+j_2-1}+\ldots)=0.
\end{equation}
 $v$ is the smallest operator formed only by the parameters from the higher order pole coefficients which could be 
 seen from expanding the spectral curve explicitly, and it has the scaling dimension determined by (\ref{scaling}) :
\begin{equation}
[v]=k_2[x]-((k_2-1)(n_1-2)+j_2-1)[z]={(n_1-1)k_1+k_2j_1-k_1j_2\over k_1n_1-k_1+j_1}.
\end{equation}
This operator dimension must be no larger than one as it must be a coupling constant or mass, and we have 
\begin{equation}
[v]\leq 1\rightarrow k_2j_1-k_1j_2\leq j_1\rightarrow {j_1\over k_1}\leq {j_2\over k_2-1}\rightarrow {j_1\over k_1}< {j_2\over k_2}.
\end{equation} 
which is a contradiction of our assumption that ${j_1\over k_1}>{j_2\over k_2}$. 

 We are left with the only other possibilities that there are only two blocks and $k_2=1$. In this 
case, the leading order matrix reads
\begin{equation}
\Phi={z^{n+{j\over {k-1}}-2}}diag(0,1,\omega,....\omega^{(k-1)-1}).
\end{equation}

 \begin{figure}[htbp]
\begin{center}
\includegraphics[width=8cm]{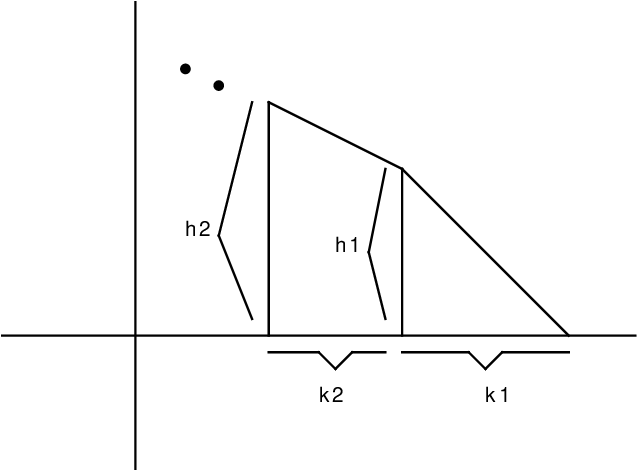}
\end{center}
\caption{The graphical form of the SW curve defined by an irregular singularity whose leading matrix has block with size $k_1, k_2, \ldots$.}
\label{proof}
\end{figure}

Therefore, the irregular singularities which could be used to define the AD theories are classified as follows:

1. Type I singularity:

\begin{equation}
\Phi=z^{n+\frac{j}{k}-2}diag(1,\omega,....\omega^{n-1}),~~0<j\leq k.
\label{canonical}
\end{equation}

2. Type II singularity:
\begin{equation}
\Phi=z^{n+{j\over k-1}-2}diag(0,1,\omega,....\omega^{(n-1)-1}), 0<j<k-1.
\label{canonical}
\end{equation}

3. Type III singularity: one could consider the degeneracy of the above irregular singularity. 
There are two scenarios one could consider:

Case 1: If the pole of the order is integer, then the singularity is specified by  a sequence of Young Tableaux 
$Y_n\subseteq Y_{n-1}....\subseteq Y_1$, in which $Y_{j-1}$ is derived by further partitioning each column 
of $Y_j$.

Case 2: If the pole of the order is fractional, one only need to determine the partition (which is determined 
by the integer points on the Newton Polygon). For instance, if the generic  singularity is 
\begin{equation}
\Phi= z^{n-2+{2\over 4}}diag(1,\omega,....\omega^{3}),
\end{equation}
there would be an integer point on the boundary of the Newton polygon. If we choose this point, the leading order 
singularity can be degenerated as the following
\begin{equation}
A_0= z^{n-2+{2\over 4}}diag(1,-1, 1, -1).
\end{equation}

Let's discuss a little bit about the geometric quantity which should be matched with the physical consideration. For
example, the number of Coulomb branch dimension should be matched with the dimension of the base
of the Hitchin fibration (half of the  dimension of Hitchin moduli space).  The 
mass parameter is basically determined by the form of the coefficient of the first order pole.

The above list exhausts all the possible irregular singularity for defining a SCFT, in the following we are 
going to study them in some detail.

\subsubsection{Type I SCFT: $(A_{k-1},A_{N-1})$ theory}
\begin{figure}[htbp]
\small
\centering
\includegraphics[width=4in]{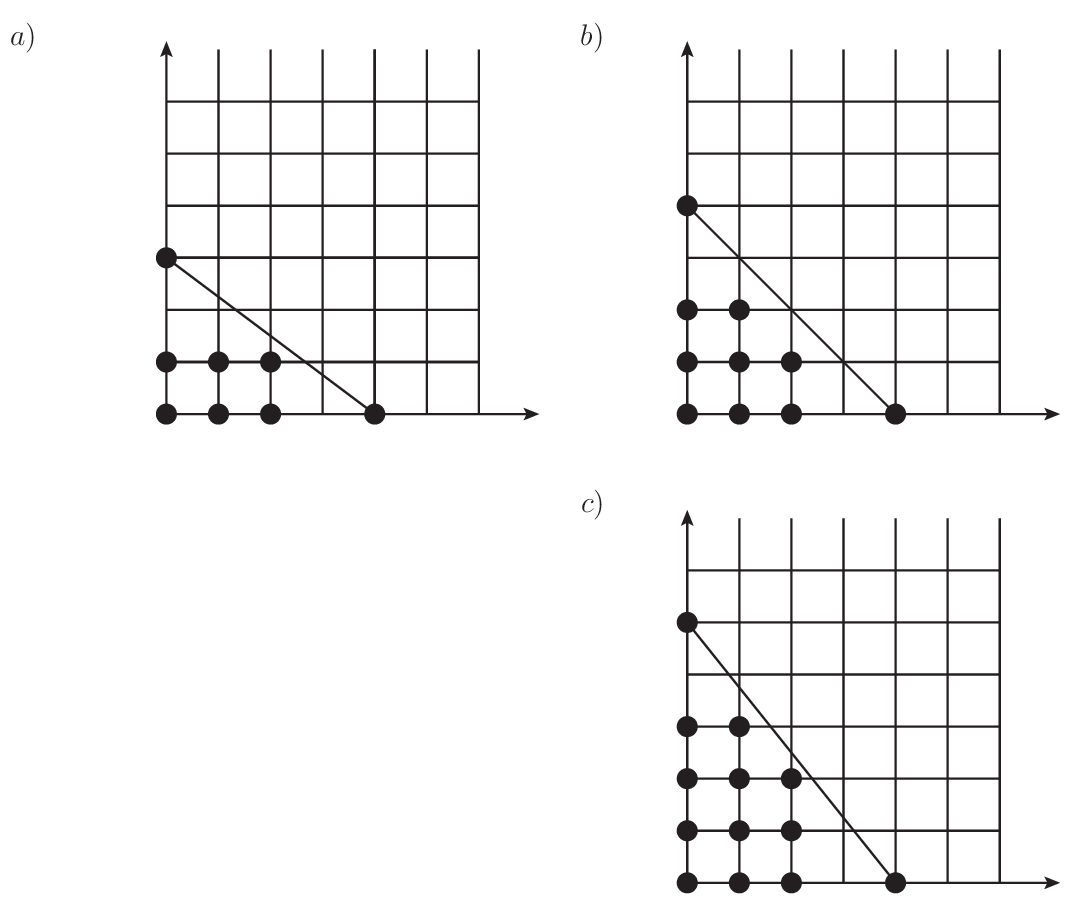}
\caption{a) Seiberg-Witten graph for $(A_3, A_2)$ theory; b) Seiberg-Witten graph for $(A_3, A_3)$ theory;
c) Seiberg-Witten Graph for $(A_3, A_4)$ theory.}
\label{SU(4)}
\end{figure}

Let's compactify six dimensional $A_{k-1}$ theory on a sphere with a
type I irregular singularity, and the Seiberg-Witten curve for the 4d SCFT at the singularity (turning on just the leading order matrices) is
\begin{equation}
x^k+z^N=0.
\end{equation}
The order of pole of this singularity is $r=N/k+2$ and the scale invariance is used to fix the coefficient of $z^N$ to be 1. The scaling dimension of $x$ and $z$ are
\begin{equation}
[x]={N\over N+k},~~[z]={k\over N+k}.
\end{equation}
This theory is called $(A_{k-1}, A_{N-1})$ theory as named in \cite{Cecotti:2010fi} because 
the BPS quiver is the product of two Dynkin diagrams of the corresponding type.  
The Seiberg-Witten curve under general deformation is very easy to find: one just 
use the integer points bounded by the Newton Polygon of the corresponding irregular 
singularity, and each integer point with coordinate $(m,n)$ 
correspond to a deformation term  $x^mz^n$.
The points on $x=k-1$ line are not used since the gauge group is 
$SU(k)$. All the points on $z=N-1$ are also eliminated. In most cases, there is only 
one point bounded by the Newton polygon on this line and the elimination can be done using
the translation invariance. In other cases, the extra deformations are not allowed for 
this particular SCFT, but they are necessary for the SCFT derived by adding 
an extra regular singularity as we discuss later.  We depict some examples in fig.~\ref{SU(4)} which are derived using six dimensional $A_3$ theory.

For example, according to our prescription, the Seiberg-Witten curve for $(A_3, A_2)$ theory is 
\begin{equation}
x^4+(v_1z+v_2)x^2+(v_3z+v_4)x+z^3+v_5z+v_6=0.
\end{equation}
One could easily find the scaling dimension of the various operators and check 
explicitly that there is a coupling constant for every relevant operator to be paired with.

The pairing between the relevant operators and coupling constants can be proven for general case. 
Let's first  analyze the points on the line $x=k-2$. The deformations on this line 
have form $u_i x^{k-2}z^i$, and the scaling dimension of $u_i$ is 
\begin{equation}
[u_i]=2[x]-i[z]. 
\end{equation}
Any non-negative integer $i$ is allowed if the above  scaling dimension is positive for that $i$.  Now let's 
consider the deformation on the line $x=0$ and the deformations have the form
$v_jz^{N-j}$ with $ 2\leq j\leq N$, the scaling dimension of $v_j$ is 
\begin{equation}
[v_j]=j[z].
\end{equation} 
The sum of the operator dimensions are
\begin{equation}
[u_i]+[v_j]=2[x]+(j-i)[z].
\end{equation}
Using the condition $2[x]+2[z]=2$, we conclude that the paring is there for all the relevant operators on
these two lines by taking $j-i=2$.
Similar analysis can be applied to the deformations on the line $x=k-2-l$ and $x=l$.
Notice that the exclusion of the operators at line $z=N-1$ is crucial for the pairing.

If the corresponding irregular singularity has integer pole, then there are $N-2$ extra parameters 
in the leading order coefficient which one could turn on. These parameters are dimensionless and 
luckily we do find $N-2$ dimension $2$ operators in the spectrum, so these dimensionless coupling 
constants are naturally identified with the exact marginal deformations.  Unlike the familiar $\mathcal{N}=4$
SYM theory, it seems that these AD theories do not have a weakly coupled description in
the conformal manifold.

It is interesting to note that the operator spectrum is completely same if we exchange 
the $x$ and $y$ axis, therefore $(A_{k-1},A_{N-1})$ theory is
equivalent to $(A_{N-1},A_{k-1})$ theory.  

By explicitly calculating the number of 
Coulomb branch deformation and the mass parameters, one can prove that the 
rank of the charge lattice of the $(A_{N-1},A_{k-1})$ theory is 
\begin{equation}
R=2n_c+n_f=(k-1)(N-1).
\end{equation}
This is in agreement with the number of quiver nodes in the BPS quiver as suggested by its name.

The above fact can be checked using the geometric property of the irregular singularity. 
Let's assume the singularity has integer pole and the order is $r={nk\over k} +2=n+2$, which actually defines the $(A_{k-1},A_{nk-1})$ theory. 
The dimension of the moduli space with such an irregular singularity is  
\begin{equation}
d=(n+2)(k^2-k)-2(k^2-1)=(nk-1)(k-1)-(k-1).
\end{equation}
This number equals to $2n_r$ where $n_r$ is the rank of gauge group at
the generic point of Coulomb branch, and there are also $n_f=k-1$ mass parameters,  so
the rank of the charge lattice from the geometric consideration is 
\begin{equation}
R=(k-1)(nk-1),
\end{equation}

\subsubsection{Type II SCFT}
\begin{figure}[htbp]
\small
\centering
\includegraphics[width=6cm]{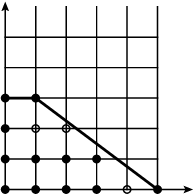}
\caption{The Seiberg-Witten curve is found from the marked integer points bounded by the Newton polygon.}
\label{type2}
\end{figure}

This type of SCFT is defined by putting a type II irregular singularity on the sphere, and the Seiberg-Witten 
curve at the SCFT point is
\begin{equation}
x^k+z^N x=0,
\end{equation}
so the scaling dimensions of the coordinates $x$ and $z$ are
\begin{equation}
[x]={N\over k-1+N},~~~[z]={k-1\over k-1+N}. 
\end{equation} 
The order of  pole of  the irregular singularity is $r=N/(k-1)+2$. Similarly, the general deformations for the theory is completely fixed by the Newton polygon of 
the corresponding irregular singularity.  See an example in figure. \ref{type2}. The $k-1$
block is the same as a $(A_{k-2}, A_{N-1})$ theory and the analysis of the spectrum on this part is the same as 
the Type I theory. One can also check that there is a coupling constant for every relevant operator 
on the extra line $x=0$.

\newpage
\subsubsection{Type III SCFT}
The SCFT defined by using type III irregular singularity is a little bit harder to analyze. Let's 
first study the integer pole case, and the singularity is specified by 
a sequence of Young Tableaux  $Y_n\subseteq Y_{n-1}\dots\subseteq Y_1$ with $n=N/k+2$.
The Seiberg-Witten curve at the singularity is 
\begin{equation}
x^k +u_1x^{N-2}+u_2x^{N-3}+\ldots z^{N}=0,
\end{equation}
where $u_i$ is tuned such that the roots have the degeneracy determined by $Y_n$. 
The  Seiberg-Witten curve under deformation is the same as the non-degenerating case, however,
not all the operators are independent and they satisfy quite complicated relations. We 
would like to know the scaling dimensions of the  spectrum which is contained in the corresponding 
Young Tableaux sequence as the regular puncture case. The general Seiberg-Witten curve is 
\begin{equation}
x^k+\sum_{i=2}^k\Phi_i(z)z^{k-i}=0
\end{equation}
The highest order of $z$ in $\Phi_i$ whose coefficient is a Coulomb branch operator is determined by 
the following data:
\begin{equation}
m_i=\sum_jp_i^{(j)}-2i+1,
\end{equation}
where $p_i^{(j)}=i-s_i$ and $s_i$ is the height of the $i$th box in the $j$th Young Tableaux, and $\Phi_i$ has the following form
\begin{equation}
\Phi_i=\ldots+u_1z^{m_i-1}+u_2z^{m_i-2}+\ldots+c,
\end{equation}
and the coefficients labeled by $u_1, u_2, \ldots, c$ are the independent Coulomb branch operators.

There are two special cases deserved further remarks, and such situations also happen for 
the theory defined  using regular punctures. The first case is when the number $m_k$ in the leading order coefficient is zero or negative, and one 
can not use the above method to find the spectrum,  currently we do not find a systematic way of 
dealing with it. The other case is that $m_k$ is non-zero, but there are some other $m_j$ which is negative, 
and one can still use the above calculus to find the spectrum, but the number of 
Coulomb branch dimension is not the same as the dimension of the base of the Hitchin fibration.

If the order of pole is fractional, one can follow the similar method to find the spectrum for the degenerating case,
we leave this to interested reader.

These configurations do not always define a SCFT since there would be no pairing between the dimensionless coupling (encoding in
the leading order matrix of the irregular singularity) and the dimension two operators. The determination of when the 
theory is a SCFT is left for future study. 

\subsubsection{Adding one more regular singularity: Type IV SCFT}
\begin{figure}[htbp]
\small
\centering
\includegraphics[width=6cm]{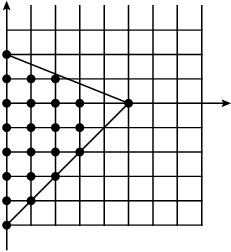}
\caption{The Seiberg-Witten graph for the theory with a irregular singularity and a regular singularity.}
\label{type4}
\end{figure}

We can add another regular singularity to above cases and find the AD theory in the IR.
One example is shown in figure. \ref{type4}. Notice
that all the possible deformations of the irregular singularity are turned on such that the relevant deformations 
from the regular singularity can have a coupling constant.  So those previous 
prohibited coupling constants do play an important role here.

The Seiberg-Witten curve is calculated using the data in Young Tableaux $Y_0$ of the regular puncture. For example, 
the independent Coulomb branch parameters due to the regular singularity has the following form in the Seiberg Witten curve
\begin{equation}
\Phi_i=\ldots+(v_{1}z^{-1}+\ldots+v_{n_i}z^{-n_i})
\end{equation}
here $\Phi_i$ is the degree $i$ differential; and $n_i=i-s_i$,  with $i$ the $i$th box and $s_i$ the height of the $i$th 
box in Young Tableaux $Y_0$. 

The regular puncture can carry
non-abelian flavor symmetry and this type of SCFT is important in constructing asymptotical 
free theories. They appear naturally in theory defined by putting multiple 
irregular singularities on the sphere: the corresponding 
4d theory is an asymptotical free theory which is formed by gauging these AD theories
together, and the irregular singularity is always at the boundary in the 
degeneration limit.

\subsection{3d Mirror symmetry}
The three dimensional mirror theory is similarly found by finding the quiver attached to an irregular singularity.
Right now,  the 3d mirror is known only for theories defined using irregular singularity with integer order of pole, which is classified by 
a sequence of Young Tableaux $Y_n\subseteq Y_{n-1}....\subseteq Y_1$, where $Y_{j}$ is derived by
further partitioning each column of $Y_{j+1}$.   The mathematical considerations appear in \cite{Boalch:2008pb} and 
the interpretation of their results are the mirror symmetry for the Argyres-Douglas theory.

\begin{figure}[htbp]
\small
\centering
\includegraphics[width=12cm]{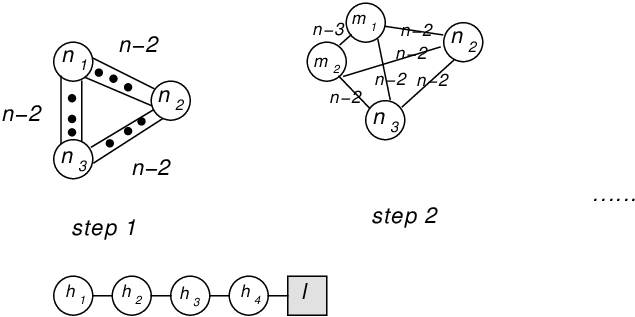} 
\caption{Step 1: If the first Young Tableaux $Y_n$ has partition $[n_1, n_2, n_3]$, then there is a quiver tail with three nodes each with rank $n_i$, there are $n-2$ arrows
between those nodes. Step 2: If  $n_1$ is further partitioned in to $[m_1, m_2]$ in $Y_{n-1}$, we split the quiver node with rank $n_1$ into two quiver nodes each 
with rank $m_1$ and $m_2$, the arrows between $m_i$ and $n_1$, $n_2$ are the same; the new arrows between $m_1$ and $m_2$ are $n-3$. The similar procedure 
is done for other Young Tableaux and we stop at $Y_2$. Bottom: If a  column with height $l$ in $Y_2$ is further split into $[l_1, l_2,\ldots,l_t]$, one attach 
a quiver tail to the node with rank $l$  in quiver  determined by $(Y_n,\ldots Y_2)$  }
\label{3dge}
\end{figure}

The three dimensional mirror is derived step by step as the following:
Assuming the partition of $Y_n$ 
is $[n_1,n_2,....n_r]$, the quiver for the first step has $r$ nodes and each node has 
rank $n_i$; there are also $n-2$ arrows representing bi-fundamental between each node, see figure. \ref{3dge} for step 1; in the second step, 
if one of $n_i$ of $Y_1$ is further partitioned as $[m_{i1}, m_{i2},\ldots m_{is}]$ in Young  Tableaux $Y_{n-1}$, the quiver 
node from step 1 with rank $n_i$ is split into $s$ quiver nodes whose rank is determined by the partition $m_{ij}$; the quiver arrows between 
these new quiver nodes are $n-3$, furthermore, we keep $n-2$ arrows between all the split nodes and the nodes in other clusters, see figure. \ref{3dge} for step 2.
One do the similar splitting for each Young Tableaux until  $Y_2$, and  get a quiver  with several nodes and arrows 
between them. Notice that the sum of the total rank  of all the quiver nodes are $k$.

The special treatment is needed for $Y_1$ or the mass matrices:  if  one of the column of $Y_2$ has height $l$ and 
this one is further partitioned as $[l_1,l_2,\ldots ,l_t]$ in $Y_1$,  we do not decompose this node as have been done before;
instead,  we attach a quiver leg as done for the regular singularity with 
total boxes $l$ \cite{Benini:2010uu}. More specifically, define $h_i=\sum_{t}^{t-i+1} l_j$, the quiver tail for this part is shown in figure. \ref{3dge}.

If there is an extra regular singularity specified by a Young Tableaux, then one first attach a quiver tail as described in the bottom of  figure.\ref{3dge}, with total boxes $k$.
Then spray the $U(k)$ node as the pattern determined by the Young Tableaux $Y_2$ of the  the irregular singularity, 
and finally we glue the quiver of irregular singularity and regular singularity by identifying the sprayed nodes of regular singularity tail.

Let's provide some explicit examples. 

\begin{figure}[htbp]
\small
\centering
\includegraphics[width=12cm]{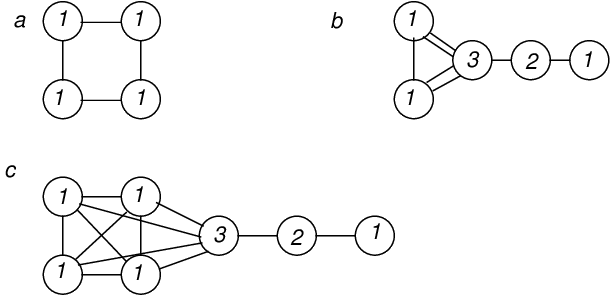}
\caption{The 3d mirror for the AD theory described in Example 1, 2, 3.}
\label{3dmr}
\end{figure}

\textbf{Example 1}: There is an order 3 irregular singularity with Young Tableaux $[2,2]$,  $[1,1,1,1]$, $[1,1,1,1]$, the quiver  is shown in 
figure. \ref{3dmr}a.

\textbf{Example 2}: There is an order 4 irregular singularity with Young Tableaux $[3,2]$,  $[3,1,1]$, $[3,1,1]$, $[1,1,1,1,1]$, the quiver  is shown in 
figure. \ref{3dmr}b.

\textbf{Example 3}: There is an order 3 irregular singularity with Young Tableaux $[1,1,1,1]$, $[1,1,1,1]$, $[1,1,1,1]$ and a regular full singularity, 
see figure. \ref{3dmr}c for the quiver.

\newpage
\subsection{Equivalence between SCFTs}
\begin{figure}[htbp]
\small
\centering
\includegraphics[width=10cm]{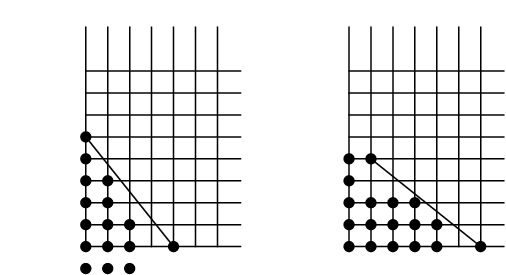}
\caption{a) A Type IV SCFT which is defined from six dimensional $A_3$ theory with an irregular singularity whose form is given by the 
Newton polygon, and there is an extra simple regular singularity. We only write the Coulomb branch parameters from the regular singularity. b) A
Type II SCFT which is defined from $A_5$ theory with proper irregular singularity, and this theory is equivalent to (a) by calculating the spectrum explicitly.}.
\label{iso}
\end{figure}

$(a)$ We have already shown that the $(A_{k-1}, A_{N-1})$ theory is identical to $(A_{N-1},A_{k-1})$ theory. There is another isomorphism
between Type II and Type IV theories. Consider the Type IV theory with a minimal regular singularity,  and the irregular singularity
is the one  defining $(A_{k-1}, A_{N-1})$ theory, we need to take $N\geq k$ here. One example 
is shown in figure. \ref{iso}a. The scaling dimensions of $[x]$ and $[z]$ is still
\begin{equation}
[x]={N\over N+k},~~~~[z]={k\over N+k}.
\end{equation}
There is one extra mass parameter from the regular puncture and those extra $(k-1)$ Coulomb branch operators have dimension
\begin{equation}
u_i=[z]+i[x],~~~2\leq i\leq k.
\end{equation}

Now let's consider a Type II theory shown in figure fig.~\ref{iso}(b) which is realized  using six dimensional $A_{N}$ theory. The 
triangle part of the Newton polygon is the same as $(A_{N-1}, A_{k-1})$ theory which is equivalent to $(A_{k-1}, A_{N-1})$ theory.  The 
scaling dimension is 
\begin{equation}
[x^{'}]={k\over N+k},~~~[z^{'}]={N\over N+k}.
\end{equation}
We just need to find the scaling dimension of the extra operators on the line $x=0$, and they take the form 
\begin{equation}
v_i=[x^{'}]+i[z^{'}],~~~0\leq i\leq k.
\end{equation}
So $v_0$ is a coupling constant, $v_1$ is  a mass parameter. The Coulomb branch  spectrum are exactly 
the same with the above realization by noting that $[x]$ and $[z^{'}]$ have the same scaling dimension. This isomorphism is achieved by exchanging $x$ and $z$ coordinates, which
is also the mechanism for the isomorphism of  the $(A,A)$ theory.

Let's make some further comments for $k=2$:  the first realization using the regular singularity is the familiar $(A_1, D_{N+2})$ theory which is given by using six dimensional $A_1$ theory.
We now show that they can also be realized by using just one irregular singularity of  $A_{N}$ theory.

There are more similar isomorphisms and the interested reader can have some fun in identifying them.
\subsubsection{Irregular representations for theories of class ${\cal S}$} 
\begin{figure}[htbp]
\small
\centering
\includegraphics[width=12cm]{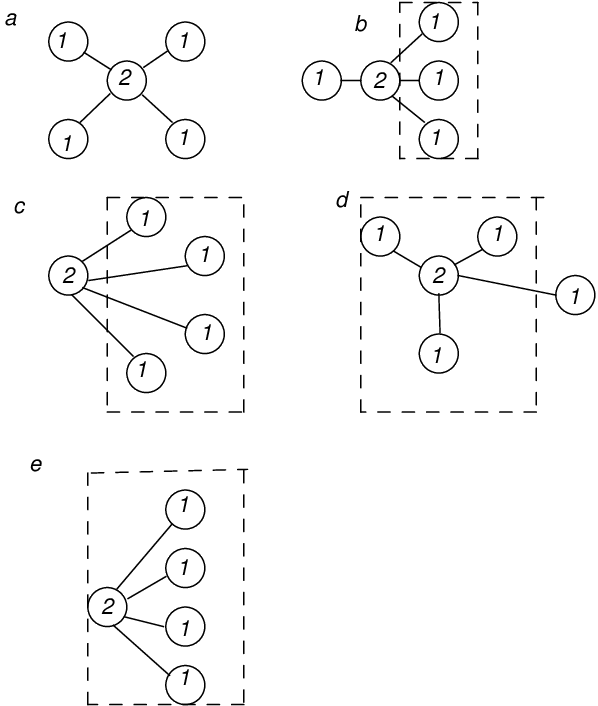}
\caption{Different decompositions of the 3d mirror of SU(2) theory with four flavor. Each decomposition corresponds to a 
combination of singularity structure. a: there are four regular singularities on the sphere, each one has a quiver tail $1-2$ and 
the 3d mirror is derived by gluing the $U(2)$ flavor symmetry together. b: This is a rank 3 realization, there is a regular singularity with 
quiver tail $1-2-3$ and the rank 2 irregular singularity gives a quiver with three isolated $U(1)$ nodes. So we spray the $U(3)$ node of the regular singularity quiver into 
three $U(1)$ nodes and glue it to the irregular singularity quiver. c: This is a rank four realization with a regular singularity   whose Young Tableaux is $[2,2]$, 
and the order 2 irregular singularity which  has a quiver with four isolated nodes. d: Rank 5 realization with only one irregular singularity. e. Rank 
6 realization with one  irregular singularity.}
\label{4f}
\end{figure}

There is another type of possible isomorphism coming from looking at the 3d mirror symmetry. Perhaps the 
the most surprising discovery is that all the theories defined using the sphere with regular punctures can be 
realized using the irregular singularity. Moreover, the realization is not unique.
Let's study the $SU(2)$ with four flavors in some detail and then give a full story later.  This theory has 
a six dimensional construction found by Gaiotto: it is a sphere with four regular punctures of $A_1$ theory. 
The three dimensional mirror is quite simple and is depicted in figure. \ref{4f}a. 

This same theory has lots of irregular realizations.  First, let's look at the rank  $3$ realization by
looking at the three dimensional mirror in the way depicted in figure. \ref{4f}b. This view means that we have a full regular puncture and 
an irregular puncture with two full Young Tableaux. If we arrange the mirror as the in figure. \ref{4f}c, then this is a rank $4$ 
realization: One has a regular singularity with partition $[2,2]$ and an irregular singularity with two full Young Tableaux. 
Rank $5$ representation is found by decomposing the mirror as in figure. \ref{4f}d: There is just one irregular singularity with three 
Young Tableaux whose partitions are $Y_3=[3,2],~Y_2=[2,1,1,1],~Y_3=[1,1,1,1,1] $. Finally, there can be a rank $6$ representation such 
that the only irregular singularity has partitions $Y_3=[4,2],~Y_2=[2,1,1,1,1],~Y_3=[2,1,1,1,1] $, and the 3d mirror decompostion is shown in figure. \ref{4f}e.  

In fact, there is another irregular singularity construction whose three dimensional mirror is not known. Our finding is based on 
the linear quiver $SU(2)-SU(4)$, and here the $SU(4)$ group is not conformal. It is shown in \cite{Nanopoulos:2010zb} that the six dimensional construction 
requires two irregular singularities on the Riemann sphere, and the tail $SU(2)-4$ is described by an irregular singularity with 
the structure
\begin{equation}
\Phi=z^{-1-{1\over2}}diag(1,w,-1,-w),
\end{equation}
and a $SU(4)$ full regular singularity. We conjecture this is another rank 4 realization of the $SU(2)$ theory with four flavors.

It is not hard to generalize  this to any theories in class ${\cal S}$ defined on a sphere. The 3d mirror is just a star-shaped quiver which is derived by 
gluing quiver legs from different punctures. The same 3d mirror quiver can be looked from other ways which can be realized by the irregular singularity.
There are many different realizations though. 

There are some explicit checks one could make: since they have the same 3d mirror, then their Coulomb branch dimension and 
the Higgs branch dimension are the same, and the flavor symmetries are also the same. These evidence strongly indicate that
the theory defined using the irregular singularities are the same as the theory of class ${\cal S}$ defined on a sphere with regular singularities.  Moreover,
one can find enough dimensionless gauge coupling constant by simply taking some ratios of the dimensional parameters. 
 We believe this remarkable 
duality is very important and will carry a further study in the future.

\newpage
\subsection{A conjecture for $R(B)$ and central charges $a$, $c$}
The calculation of the central charges are quite similar as what has been done for $A_1$ and $A_2$ theory:
 first use the mirror symmetry and the scaling dimension of the operator spectrum to calculate the central charge for 
 a subset of theories, then plug the results into  the formula (\ref{central}) to get the function $R(B)$; Finally, use the $R(B)$ 
 to calculate central charges for other theories in this class. We assume that a 
 uniform formula for $R(B)$ exists and an explicit form will be given for  $(A_{k-1},A_{N-1})$ theory.
Our conjecture is based on following known formula for $A_1$ and $A_2$, and
 the invariance under the exchange of $k$ and $N$. Our previous calculations on $A_1$ and $A_2$ theory gave
\begin{equation}
A_1:~~~R(B)={1\over2}{N(N-1)\over(N+2)}
\end{equation}
\begin{equation}
A_2:~~~R(B)={3\over 2}{N(N-1)\over (N+3)}
\end{equation}

The general formula we conjecture is 
\begin{equation}
A_{k-1}:~~~~R(B)={k(k-1)\over 4}{N(N-1)\over (N+k)},
\end{equation}
which is invariant under changing $k$ and $N$ and recover the $A_1$ and $A_2$ case.

\subsubsection{Confirmation from mirror symmetry}
We may check the conjecture using three dimensional mirror. Let's consider $(A_{k-1},A_{nk-1})$
theory whose three dimensional mirror is known. Let's first calculate its central charge using the above conjectural form of $R(B)$. Since the
Seiberg-Witten curve is
\begin{equation}
x^k+(u_1z^{2n-1}+\ldots+u_{2n})x^{k-2}+(v_1z^{3n-1}+\ldots+v_{3n})x^{k-3}+\ldots+(z^{nk}+\ldots+c_{nk})=0,
\end{equation}
then the scaling dimension of $z$ is $[z]={1\over n+1}$, and  $R(A)$ is found from the 
scaling dimension of the spectrum:
\begin{eqnarray}
R(A)=\sum(D(u_i)-1)={3 k n - 3 k^2 n + k n^2 - 3 k^2 n^2 + 2 k^3 n^2)\over(12 (1 + n)}, 
\end{eqnarray}
and $R(B)$ is found from our ansatz:
\begin{equation}
R(B)={k(k-1)\over 4}{n(nk-1)\over (n+1)}.
\end{equation}

Using the formula \ref{central}, the central charges turn out to be:
\begin{eqnarray}
c={(-1 + k) (-2 - 2 n + k n^2 + k^2n^2)\over12 (1 + n)}, \nonumber\\
a={(-1 + k) (-5 - 5 n + 2 k n^2 + 2k^2n^2)\over24 (1 + n)}.
\end{eqnarray}

Now let's calculate this using the alternative equations and 3d mirror symmetry. Since 
the Coulomb branch dimension of the 3d mirror theory is $k-1$, we have
\begin{eqnarray}
2a-c={1\over 4}\sum_{i=1}^r(2D(u_i)-1)={3 - 3 k + 3 n - 3 k n - k n^2 + k^3 n^2\over12 (1 + n)}\nonumber\\
a-c=-{k-1\over 24}. 
\end{eqnarray}
We also find
\begin{eqnarray}
a={(-1 + k) (-5 - 5 n + 2 k n^2+2k^2n^2)\over24 (1 + n)}, \nonumber\\
c={(-1 + k) (-2 - 2 n + k n^2+k^2n^2)\over12 (1 + n)},
\end{eqnarray}
which is the same as the earlier results using $R(B)$.  In the large $N$ and large $k$ limit, we have 
\begin{equation}
{a\over c}=1,~~~~~a={1\over 12} k^2N.
\end{equation}
So it is possible to find the gravity dual for this class of theory in the large $N$ and large $k$ limit.  
Once we know $R(B)$, it is easy to find the central charges for the other $(A_{k-1},A_{N-1})$ theory where
3d mirror theory is not known yet.  The same method can be used to calculate $R(B)$ of other  class of SCFTs if the 3d mirror theories
for a  subclass can be found.

\subsection{AD points from $N=2$ QCD}

We want to identify linear quiver gauge theory whose Coulomb branch contains the specific
AD points. The method is to consider the collision of singularities: The collision is possible if 
the two singularities with fractional order have the same leading behavior, i. e. the leading order matrix is the same. 

In the case of $SU(k)$  QCD with $N_f$ flavors,  the six dimensional realization is not unique and 
labeled by a partition of $N_f=n_1+n_2$ with $n_1\leq n_2\leq k$, which basically means splitting 
the flavors into two groups.  Each group of the flavors is described by an irregular singularity.
The irregular singularity for $l< (k-1)$ flavors are
\begin{eqnarray}
N_f=l:~~\Phi={1\over z^{1+1/(k-l)}}(0,\ldots 0,1,\omega,...\omega^{k-l-1})+...
\end{eqnarray}
For $N_f=k-1$, the irregular singularity has two Young Tableaux: $Y_2$ is 
a simple one and $Y_1$ is a full one.
For $N_f=k$,  a full regular singularity and a simple 
regular singularity are needed. Generically, the collision is possible when there are even number of  flavors which are divided into two equal parts, 
but the resulting irregular singularity may not define an AD theory.  According to our above classification of the AD theories and the collision rule, we can find 
AD theories in the following scenarios:

a. When $N_f=0$, the colliding irregular singularity  has the following form
\begin{equation}
\Phi={1\over z^{2+2/k}}(1,\omega,...\omega^{k-1}).
\end{equation}
The AD theories defined by this irregular singularity is the $(A_{k-1}, A_{1})$ theory which is in agreement with
the result found in \cite{Eguchi:1996vu}.

b. When $N_f=2$, the colliding irregular singularity has the following form
\begin{equation}
\Phi={1\over z^{2+2/(k-1)}}(0,1,\omega,...\omega^{k-2}).
\end{equation}
The AD theories defined by this irregular singularity is the $(A_1, D_{k+1})$ theory using the isomorphism we discovered , which is  also in agreement with
the result found in \cite{Eguchi:1996vu}.

c. When $N_f=k$ and the partition is $N_f=k\oplus 0$, then there is an irregular singularity describing nothing, and a full and simple regular singularities.
Therefore one can collide the irregular singularity with the full regular singularity and get the following combined one
\begin{equation}
\Phi={1\over z^{2+1/(k)}}(1,\omega,...\omega^{k-1}).
\end{equation}
There is another simple regular singularity, so it is a type IV SCFT in the collision limit.

d. When $N_f=k+1$ and the partition is $N_f=k\oplus1$, and the collision of the irregular singularity and the regular full singularity produce an irregular singularity
\begin{equation}
\Phi={1\over z^{2+1/(k-1)}}(0,1,\omega,...\omega^{k-2}).
\end{equation}
This is also a type IV SCFT since there is an extra simple regular singularity.

Generically, the above AD theories are all we find from $SU(k)$ QCD.

\subsection{General irregular singularities}
What is the interpretation of the other irregular singularities?  If the leading order block has rank $r>2$,  then $z$
has non-trivial scaling dimension, and
our conjecture is that all the other blocks should be deformed such that they have the same rank as the first block, 
therefore we have a AD theory. So we really need not to consider those general irregular singularities if our 
conjecture is true.

The case is quite different if the leading order block has rank $r\leq 2$, in this 
case, $z$ has to transform trivially under the rotation and every block is in the same footing (we have used this fact in our previous examples). 
It is shown that some of the irregular singularities represent the 
asymptotical free theory \cite{Nanopoulos:2010zb}. It is interesting to study these cases in more detail and 
we leave it to the future.

\section{Six dimensional representation of known examples}
We have already shown that  $(A_1, A_N)$, $(A_1,D_N)$ and $(A_1, E_N)$ theory  can be 
engineered  using six dimensional $A_1$ and $A_2$ theory with appropriate irregular singularities.
More generally, the $(A_{k-1}, A_{N-1})$ theories can also be engineered using the six dimensional $A_{k-1}$ theory
or $A_{N-1}$ theory. Let's now give more examples.

There are six dimensional  rank $N$  generalizations for $A_0,A_1,A_2,  E_6, E_7, E_8$ theory (the name is 
coming from the flavor symmetry and Kodaira's classification for singular fibre). 
Rank N $E_6$, $E_7$ and $E_8$ theory can be realized by compactifying six dimensional $A_{3N-1}$,
$A_{4N-1}$ and $A_{6N-1}$ theory on a three punctured sphere with just regular singularities \cite{Benini:2009gi}. 

We now use irregular singularity to engineer rank $N$  $A_0, A_1, A_2$ theories.
We compactify $A_{2N-1}$ theory on a sphere with the following irregular singularity
\begin{equation}
A_0: \Phi={1\over z^{4-1/2}}diag(1,1,...1,-1,-1,...-1)+...
\end{equation}
The leading order coefficient has N roots 1 and N roots -1, and the sub-leading order term has the same pattern.
The operators have dimension ${6\over 5}, 2\times{6\over 5},\ldots {6\over 5}N$.

Rank N $A_1$ theory is engineered using $A_{2N-1}$ theory and the following irregular singularity
\begin{equation}
A_1:\Phi={1\over z^{4}}diag(1,1,...1,-1,-1,...-1)+...
\end{equation}
where the coefficient of ${1\over z^3}...{1\over z}$ have the same type of matrix as the leading order. The
spectrum has scaling dimension ${4\over 3},{8\over 3},...{4\over 3}N$, and one has a mass parameter.

Two singularities are needed for rank $N$ $A_2$ theories: One is irregular whose form is
\begin{equation}
A_2:\Phi={1\over z^{3}}diag(1,1,...1,-1,-1,...-1)+...
\end{equation}
where the coefficient of ${1\over z^2}...{1\over z}$ have the same type of matrix as the leading order,
and the other one  is a regular singularity with Young Tableaux $[N,N]$. The theory has $N$ Coulomb
branch operators with dimension ${3\over 2},{3\over 2}\times 2, \ldots, {3\over2}\times N$, and  two mass parameters.

\begin{figure}[htbp]
\small
\centering
\includegraphics[width=2.0in]{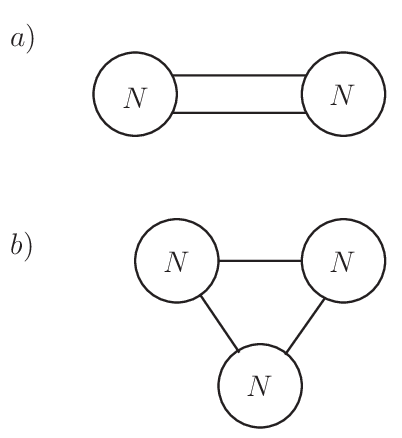}
\caption{(a) 3d Mirror for rank N $A_1$ theory. (b) 3d Mirror for rank N $A_2$ theory. }
\label{rank-N}
\end{figure}

The mirror theory for rank N $A_1$ and $A_2$ is depicted in fig.~\ref{rank-N}.
One can calculate the central charges $a$ and $c$ using the information of spectrum
and mirror symmetry. For Rank N $A_1$ theory, the Coulomb branch dimension of the mirror is $2N-1$, we have
\begin{eqnarray}
2a-c={1\over 4}\sum_i(2D(u_i)-1)={1\over12} (N + 4 N^2), \nonumber\\
a-c=-{2N-1\over 24},
\end{eqnarray}
so
\begin{equation}
a={1\over 24} (-1 + 4 N + 8 N^2),~~c={1\over 12} (-1 + 3 N + 4 N^2).
\end{equation}
Similarly, for rank N $A_2$ theory, the mirror Coulomb branch dimension is $3N-1$, we have
\begin{eqnarray}
2a-c={1\over 8} (N + 3 N^2),\nonumber\\
a-c=-{3N-1\over 24}.
\end{eqnarray}
and
\begin{equation}
a={1\over 24} (-1 + 6 N + 9 N^2),~~c={1\over 24} (-2 + 9 N + 9 N^2).
\end{equation}
These results are in agreement with \cite{Aharony:2007dj}. 

Some of the SCFTs found using singularity theory in \cite{Cecotti:2011gu,DelZotto:2011an} can also be engineered using 
the irregular singularities. Let's just give one example called $W_{13}$ theory whose Seiberg-Witten curve at the singularity is 
\begin{equation}
x^4 +xy^4=0,
\end{equation}
it is obvious that this is a type II SCFT constructed using  six dimensional $A_3$ theory.

\section{Conclusion}
We have constructed a large class of AD type theories using six dimensional 
construction in this paper.  The Seiberg-Witten curve and the scaling dimensions of 
the spectrum are worked out; the central charges $a$ and $c$ are also calculated using 
3d mirror symmetry. We also identify the AD points of $SU(N)$ QCD.  

Originally, the $\mathcal{N}=2$ SCFT with Lagrangian descriptions are classified by 
ADE Dynkin diagram \cite{Katz:1997eq}. Gaiotto found a large class of new  SCFTs using regular 
punctures on Riemann surface \cite{Gaiotto:2009we}, and generically those theories do not have a Lagrangian 
description. Such theories are the generalized quiver gauge theories of the $A_n$ type with strongly coupled matter systems in the sphere case.
These results greatly increase the number of  $\mathcal{N}=2$ SCFTs.
The above theories all have the integer scaling dimensions and dimensionless 
coupling constants.  By allowing the fractional scaling dimension, we have shown that the landscape of $\mathcal{N}=2$
SCFT theory is greatly enlarged and in fact contain those previous constructed theories (at least
those defined on sphere) as a  small corner. Given the importance of the SCFT,  these huge amount of 
examples should be very helpful for us  to learn more about dynamics of quantum field theory.

Although AD theories were discovered in 1995, their properties are poorly understood and 
not many studies have been done on those theories. Given the six dimensional construction, we
believe now one can understand these theories in more detail. 
Basically, what has been done for theory of class ${\cal S}$  can also 
be extended to these AD theories with some efforts, we list some of the interesting question below:

a. We only consider six dimensional $A_n$ theory in this paper,
 It should be straightforward to generalize to six dimensional $D_n$ and $E_n$ 
theories. Some irregular singularities and the generalized monodromy for the general reductive group
are studied in \cite{MR1904670}, and one may use these results to construct new type of Argyres-Douglas theories. 

b. The general ansatz for  gravity dual of 4d  $\mathcal{N}=2$ theory is written down in \cite{Lin:2004nb}, and the gravity
solution is reduced to solving  the 3d Toda equation. The regular singularity cases are studied in \cite{Gaiotto:2009gz}.
It should be possible to find the gravity dual of the AD theories constructed in this paper by studying the irregular solution of Toda's equation. 
 Since theory of class ${\cal S}$  defined on a sphere has an irregular singularity representation, it might be easier 
to get the gravity dual using irregular singularity. The gravity dual of the higher rank 
$(A_0, A_1, A_2)$ AD theories are found in \cite{Aharony:1998xz} using F theory. Since we have given 
the explicit six dimensional construction, the above results might be useful for us to get 
the general solution.

c. The cluster coordinates for theory defined using only regular punctures are constructed in \cite{Xie:2012dw}.
That construction can also be applied to the AD theories constructed in this paper. The idea is that the irregular singularity 
creates a boundary with multiple marked points, so the geometry is really a disk with marked points and the construction of 
\cite{Xie:2012dw} can be easily applied here. The problem 
reduces to  identify how many marked points and what kind of Young Tableaux should 
be put there \cite{Xie:2012jd} for a specific irregular singularity.  Once the cluster
coordinates are found, one can use them to study classification of line operators, 
BPS spectrum and wall crossing, surface operators, etc.

d. The AGT \cite{Alday:2009aq} conjecture for the $A_1$ AD theories are studied in \cite{Gaiotto:2012sf, Bonelli:2011aa}, which involves the irregular 
conformal block of the Liouville theory. It should be possible 
to generalize the ACT conjecture to the higher rank AD theory constructed in this paper. Since these theories do not have 
a Lagrangian, one need to use some other methods to find the Nekrasov partition function, the 
method described in \cite{Nekrasov:2010ka} should be helpful if we have good coordinates for the moduli space of  flat connection in the presence of irregular singularity. 
The AGT conjecture for asymptotical free theories constructed from 
$A_1$ theory is also studied \cite{Gaiotto:2009ma} and it is also interesting to extend that construction to all the asymptotical
free theories constructed in this paper.

e. The Seiberg-Witten solution defines an integrable system, and the Nekrasov-Shatashvili relation between the partition function of the 4d $\mathcal{N}=2$ theory 
and the Yang-Yang function of the integrable system can be generalized here. It seems that the  Darboux coordinates  constructed  in \cite{Nekrasov:2011bc}
can be generalized to our case once the cluster coordinates are known. In fact, in the regular puncture
case, they use the loops in the pants decomposition which in fact label the lamination space for the Teichmuller
space. The generalization of lamination to the disk is described in \cite{fock-2007} and the description of the 
corresponding expression for the Darboux coordinates should be possible once we know the 
cluster coordinate.

f. Mathematically, irregular singularity and the integrable system have been studied extensively 
recently \cite{feigin-2006}, we expect those results are useful for calculating the conformal block of the AD theories.
 We have shown that a theory can be realized using either regular singularity or irregular singularity,  
 would this fact  be useful for studying the tame ramification and wild ramification of Geometric 
Langlands program? Moreover, the isomonodromy deformation equations for the two realizations 
are shown to be equivalent \cite{boalch-2011}, it is interesting to see what this means for the corresponding field theory. 

g. Once we know the BPS spectrum of these higher rank AD theories, one may follow 
the approach developed in \cite{Cecotti:2011iy} to study the higher rank  3d tetrahedron theory and the corresponding 
$\mathcal{N}=2$ theory. For example, It seems that 
the BPS spectrum for the  higher rank pentagon theory can be relatively easy
to calculate (i.e. $E_7$ theory from the $A_2$ realization has 5 marked points on the boundary).

h: We have calculated the central charges $a$ and $c$ for some of the AD theories. It 
is not hard to find the RG flow between these theories and it should be fun to 
check  the $a$ theorem explicitly. General proof of the $a$ theorem in 4d is discussed in \cite{Komargodski:2011vj},
it  should be helpful to understand the proof if we can construct many explicit  examples .

i: For each AD theory, its cluster coordinates correspond to a planar bipartite graph on a 
disk. Such planar bipartite graph has been used extensively in studying the scattering amplitude
of $\mathcal{N}=4$ SYM theory \cite{ArkaniHamed:2012nw}. This same exact mathematical structure appears in two seemingly
completely different physical system. It is interesting to understand if there are any underlying 
links here. Is it possible that the physical observables of the AD theory are related to the scattering amplitude of $\mathcal{N}=4$ SYM theory?
Moreover, the planar bipartite graph has been completely classified in \cite{postnikov-2006}, can 
we use this result to give an alternative classification of  the AD theories? 
 
j:  There are some other physical quantities which one would like to calculate. For example:
It would be interesting to calculate the central charge $k$ for the type IV AD theories which 
measures the contribution to beta function when the non-abelian flavor symmetry is gauged.
It is also interesting to calculate the superconformal index for these theories, see the results for
the regular puncture case \cite{Gadde:2011uv}.  We would also like to study the Higgs branch 
of the AD theories in more detail, which seems to be captured by the Coulomb branch of the 
3d mirror theory, etc.

\begin{flushleft}
\textbf{Acknowledgments}
\end{flushleft}
We thank Nima Arkani-Hamed, Davide Gaiotto, Yu-tin Huang, Hai Lin, Juan Maldacena, Andrew Neitzke, Edward Witten,  
Yuji Tachikawa, and Peng Zhao for useful discussions. The research of DX is supported by Zurich Financial services membership and 
acknowledges support by the U.S. Department of Energy, grant DE-FG02-90ER40542.

\bibliographystyle{utphys}
   \bibliography{PLforRS}       

\end{document}